\documentclass[10pt,journal,compsoc]{IEEEtran}
\usepackage{amsmath,amsfonts}
\usepackage{mathrsfs}
\usepackage{algorithmic}
\usepackage{algorithm}
\usepackage{array}
\usepackage[caption=false,font=normalsize,labelfont=sf,textfont=sf]{subfig}
\usepackage{textcomp}
\usepackage{stfloats}
\usepackage{url}
\usepackage{verbatim}
\usepackage{graphicx}
\usepackage{threeparttable}
\usepackage{multirow}
\usepackage{multicol}
\usepackage{cite}
\usepackage{graphicx}
\usepackage{amssymb}
\usepackage{float}
\usepackage{booktabs}
\usepackage{hyperref}
\hypersetup{
    colorlinks=true,
    urlcolor=black,
}
\begin{document}

\title{WiOpen: A Robust Wi-Fi-based Open-set Gesture Recognition Framework}

\author{Xiang~Zhang~\IEEEmembership{Member,~IEEE,}
        Jingyang~Huang$^*$~\IEEEmembership{Member,~IEEE,}
        Huan~Yan,
        Peng~Zhao,
        Guohang~Zhuang,
        Zhi~Liu~\IEEEmembership{Senior Member,~IEEE},
        and Bin~Liu$^*$

\thanks{Xiang Zhang, and Bin Liu, are with CAS Key Laboratory of Electromagnetic Space Information, University of Science and Technology of China, Hefei, 230026, China.}
\thanks{Xiang Zhang, Jinyang Huang, Guohang Zhuang, and Peng Zhao, School of Computer and Information, Hefei University of Technology, Hefei, 230601, China.}
\thanks{Huan Yan, School of Big Data and Computer Science, Guizhou Normal University, Guiyang, 550025, China.}
\thanks{Zhi Liu, Department of Computer and Network Engineering, The University of Electro-Communications, Tokyo, 1828585, Japan.}
\thanks{$^*$ Corresponding authors.}}

\markboth{Journal of \LaTeX\ Class Files,~Vol.~14, No.~8, August~2023}%
{Shell \MakeLowercase{\textit{et al.}}: A Sample Article Using IEEEtran.cls for IEEE Journals}


\IEEEtitleabstractindextext{%
\begin{abstract}
Recent years have witnessed a growing interest in Wi-Fi-based gesture recognition. However, existing works have predominantly focused on closed-set paradigms, where all testing gestures are predefined during training. This poses a significant challenge in real-world applications, as unseen gestures might be misclassified as known classes during testing. To address this issue, we propose WiOpen, a robust Wi-Fi-based Open-Set Gesture Recognition (OSGR) framework. Implementing OSGR requires addressing challenges caused by the unique uncertainty in Wi-Fi sensing. This uncertainty, resulting from noise and domains, leads to widely scattered and irregular data distributions in collected Wi-Fi sensing data. Consequently, data ambiguity between classes and challenges in defining appropriate decision boundaries to identify unknowns arise. To tackle these challenges, WiOpen adopts a two-fold approach to eliminate uncertainty and define precise decision boundaries. Initially, it addresses uncertainty induced by noise during data preprocessing by utilizing the CSI ratio. Next, it designs the OSGR network based on an uncertainty quantification method. Throughout the learning process, this network effectively mitigates uncertainty stemming from domains. Ultimately, the network leverages relationships among samples' neighbors to dynamically define open-set decision boundaries, successfully realizing OSGR. Comprehensive experiments on publicly accessible datasets confirm WiOpen's effectiveness. Notably, WiOpen also demonstrates superiority in cross-domain tasks when compared to state-of-the-art approaches.

\end{abstract}

}

\maketitle

\IEEEdisplaynontitleabstractindextext

\IEEEpeerreviewmaketitle

\section{Introduction}

\IEEEPARstart{W}{i-Fi} based gesture recognition \cite{gu2022wigrunt} has garnered significant attention in recent years due to its advantages in terms of ubiquitous deployment and non-intrusive sensing. However, current studies in the field all rely on a closed-set assumption \cite{zhang2021widar3,li2020wihf,liu2023wisr}, \textit{i.e.}, each test sample is assumed to always belong to one of the pre-defined set of gesture classes. Although this conventional presumption often proves untenable in practical applications, as gesture recognition systems can invariably encounter unseen gesture classes or even non-gestural activities, close-set techniques tend to force unknown class samples to be classiﬁed into one of the known gesture classes. This limitation not only results in an poor user experience but also undermines the practicability and reliability of Wi-Fi gesture recognition systems. Therefore, it is imperative to address this drawback and develop more robust and flexible open-set gesture recognition (OSGR) approaches that can handle open-set scenarios effectively. Such methods should properly classify unknown-class samples as “unknown” and known-class samples as one of the known classes.

\begin{figure}[ht]
\centering
\includegraphics[width=1\columnwidth]{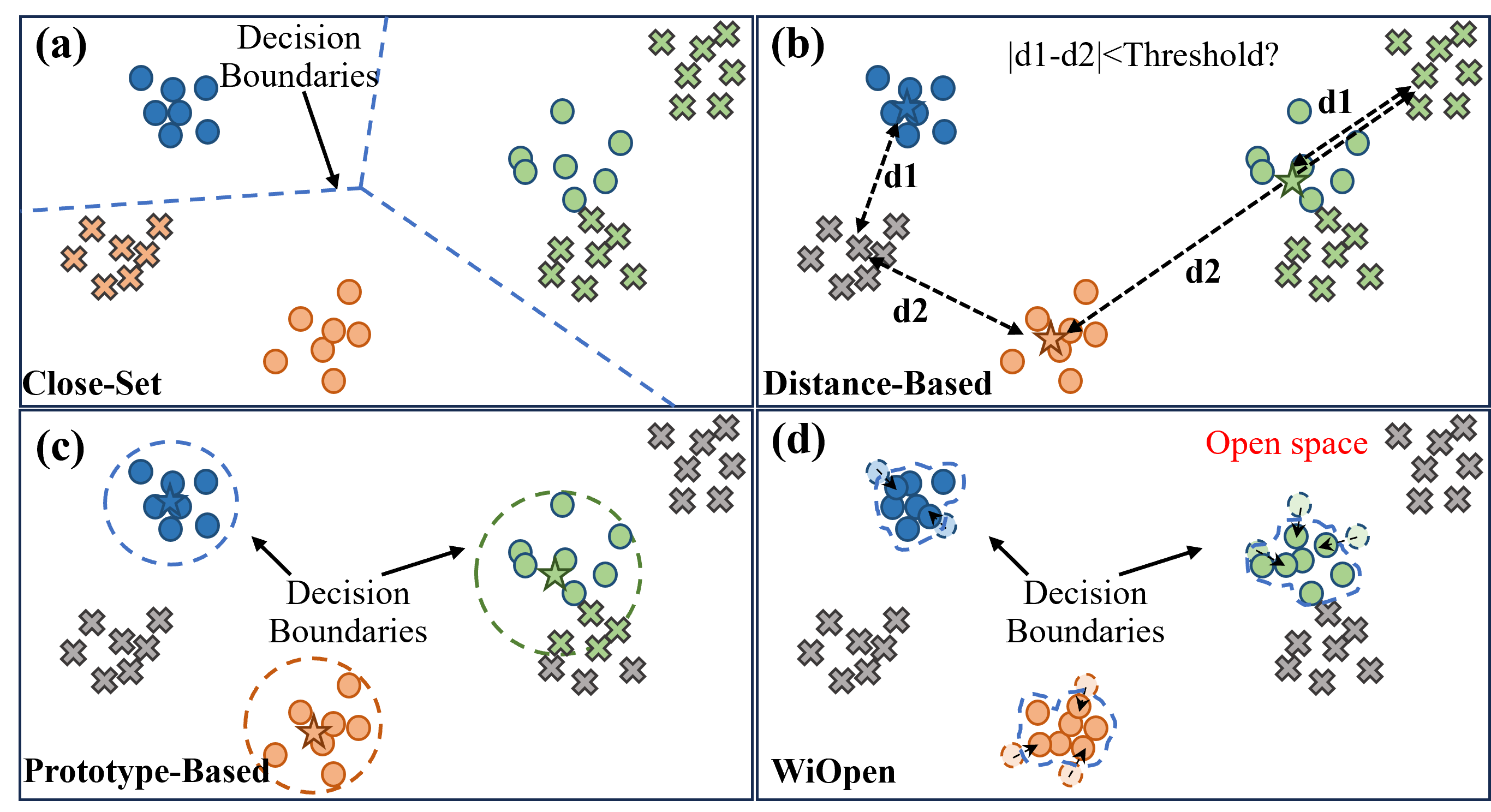}
\caption{The comparison between different approaches. (a) Close-set methods; (b) Distance-based methods; (c) Prototype-based methods; (d) Wiopen. Colored circles represent known classes samples, gray "x" symbols, colored "x" symbols, and stars represents unknown samples, misclassified samples and prototypes, respectively.}
\label{fig:intro}
\vspace{-0.2in}
\end{figure}

The objective of closed-set gesture recognition is to minimize empirical risk, which relates to the risk associated with misclassifying known classes. In contrast, OSGR not only focuses on minimizing empirical risk but also addresses open space risk. In real-world scenarios, each class's associated feature space is finite, while the area beyond these feature spaces is referred to as the open space, as depicted in Figure \ref{fig:intro}\textcolor{red}{d}. Labeling samples within the open space as known classes introduces open space risk. Traditional closed-set classifiers \cite{zhang2021widar3,liu2023wisr} usually divide the entire feature space into several known classes, and the decision boundaries of such solutions are shown in Figure \ref{fig:intro}\textcolor{red}{a}. These methods accept data from infinitely wide regions, implying that their open-space risk is unbounded, making them ineffective in identifying unknown samples. In contrast to these traditional schemes, some existing Wi-Fi-based studies have briefly touched upon the identification of unknown classes. For instance, Wione \cite{gu2021wione} constructed prototypes for all users for user identification and classified samples as unknown when their distance to the prototypes of all known classes was similar. Similarly, CAUTION \cite{wang2022caution} compared the distance between the sample and the two nearest user prototypes, classifying it as unknown when the distances were similar. Nevertheless, as illustrated in Figure \ref{fig:intro}\textcolor{red}{b}, these distance-based approaches are only effective for unknown samples in close proximity to the decision boundaries. In another examples, Tan et al. \cite{tan2020enabling} and Multitrack \cite{tan2019multitrack} developed prototypes for each activity class and employed similarity comparisons for activity recognition. Samples falling below a certain similarity threshold were labeled as unknown. While prototype-based approaches construct decision boundaries for rejecting unknowns based on distance, these boundaries cannot flexibly envelop the feature space of each class, leading to misjudgments. All of the aforementioned methods assume the existence of a prototype that can represent a class. However, due to the inherent uncertainty in Wi-Fi sensing, such an assumption is unrealistic. The uncertainty, arising from both noise and domains, gives rise to a highly scattered and irregular distribution in collected Wi-Fi sensing data. The irregularity in data distribution makes it challenging to accurately describe the data distribution using a single prototype, such as green class in Figure \ref{fig:intro}\textcolor{red}{c}.

The uncertainty caused by noise and domain are different. Noise, such as environmental and device noise, introduces random biases to the Wi-Fi sensing signals, resulting in random offsets in the data distribution and making the distribution more discrete. Domain refers to variables such as users, the location and direction of activities. According to previous studies \cite{wang2016human,niu2018fresnel}, variations in the domain can cause different mappings of activities onto Wi-Fi sensing signals, leading to directed shifts in the data. The uncertainty in Wi-Fi sensing presents two significant challenges for implementing OSGR based on Wi-Fi. The first challenge emerges from the high data dispersion caused by uncertainty, leading to confusion between classes with substantial intra-class variation and limited inter-class variation. The second challenge arises from the irregularity in data distribution, making it more complex to define decision boundaries for distinguishing unknown from known samples. 

In this paper, we present WiOpen, a robust Wi-Fi-based OSGR framework. To address the first challenge, WiOpen conducts an extensive analysis of the open-set challenges encountered in Wi-Fi gesture recognition. It introduces an uncertainty quantification method as a pivotal step. Leveraging this analysis, WiOpen takes measures to mitigate the uncertainty caused by noise during data preprocessing, achieved through the utilization of the CSI ratio. Subsequently, the OSGR network inspired by uncertainty quantification comes into play. This network learns the relationship between sensing data and Doppler frequency shift (DFS), effectively eliminating the influence stemming from static path factors. Simultaneously, it unravels the intrinsic structure of Wi-Fi data characterized by a highly irregular distribution by analyzing the relationships between neighboring samples, thereby reducing the uncertainty associated with domains. With respect to the second challenge, WiOpen suggests employing the sample's neighborhood structure rather than the structure of the sample and prototype as the criterion for decision boundary construction. The decision boundaries constructed based on distances from samples to a subset of neighbors exhibit greater flexibility compared to those constructed using distances from samples to prototypes. This approach effectively bounds open space risk. Furthermore, by designing both the training and the decision boundaries making method based on the neighborhood structure, WiOpen can address open space risk in an end-to-end manner. As illustrated in Figure \ref{fig:intro}\textcolor{red}{d}, WiOpen confines the decision boundaries of known classes, successfully mitigating open-space risk and enhancing OSGR performance.

We have implemented WiOpen and conducted a comprehensive evaluation using public datasets. The experimental results illustrate the effectiveness of WiOpen in open-set scenarios. Furthermore, when compared to state-of-the-art (SOTA) approaches, WiOpen demonstrates distinct advantages in cross-domain tasks. In summary, this paper presents the following contributions:

\begin{itemize}
\item We introduce WiOpen, the first Wi-Fi-based open-set gesture recognition system, capable of effectively reject unknown gestures while recognizing known gestures.
\item We conduct an in-depth analysis of the open-set challenges within Wi-Fi-based gesture recognition, elucidating the relationship between uncertainty and Wi-Fi OSGR performance.
\item We propose a uncertainty quantification method and an OSGR network inspired by uncertainty quantification, designed to mitigate uncertainty and achieve effective Wi-Fi OSGR.
\item We implement WiOpen and conduct extensive experiments on public datasets to evaluate its performance. Evaluations demonstrate the feasibility and effectiveness of our system.
\end{itemize}

\section{Related Works}

\subsection{Wi-Fi based Gesture Recognition}

Gesture sensing and recognition enabled by WiFi \cite{pu2013whole,wang2014eyes} can be broadly categorized into two groups: handcraft-based and deep learning-based methods. Handcraft-based approaches typically involve manual characterization of signal distortions corresponding to different gestures. In contrast, learning-based methods leverage machine learning techniques for gesture recognition. WiGest \cite{abdelnasser2015wigest} employs manual pattern construction for each gesture in received signal strength (RSS) and uses a similarity matching method for recognition. While this work is inspiring, its reliance on coarse-grained RSS indicators limits its accuracy. WiMU \cite{venkatnarayan2018multi} goes further by achieving multi-user gesture recognition using fine-grained Channel State Information (CSI). However, it relies on an exhaustive search among known gestures, leading to scalability challenges. WiDraw \cite{sun2015widraw} employs Angle-Of-Arrival (AOA) measurements for hand tracking, allowing users to draw in the air with minimal tracking error. Nevertheless, its practicality is limited by the need for over 25 WiFi transceivers around the user. QGesture \cite{yu2018qgesture} utilizes phase information for similar performance but requires knowledge of the initial hand position for tracking.

Learning-based methods shift the focus to automatic pattern recognition through data-driven techniques. This category can be further divided into shallow learning \cite{li2016wifinger, ali2015keystroke} and deep learning \cite{zhang2020wisign, zheng2019zero, li2020wihf}. Shallow learning involves training a shallow learner with handcrafted features for gesture classification. Wikey \cite{ali2015keystroke} was one of the first to explore keystroke recognition based on WiFi sensing and machine learning. WiFinger \cite{li2016wifinger} employs WiFi CSI for recognizing nine sign languages, but it requires users to be positioned in the line of sight between the transmitting and receiving antennas. While shallow learning typically requires a small training dataset, its performance is limited. Consequently, deep learning has emerged as an effective alternative. For example, WiSign \cite{zhang2020wisign} focuses on American Sign Language recognition, using amplitude and phase CSI profiles processed by a Deep Belief Network (DBN) for recognition. However, deep learning-based approaches face a critical challenge, namely, their dependence on domain-specific training.

WiDar3 \cite{zhang2021widar3} addresses this challenge by introducing a domain-independent feature, BVP, which characterizes power distribution across various velocities for cross-domain gesture recognition. WiDar3 is among the pioneers in unveiling and tackling the cross-domain issue in gesture recognition. Its meticulously crafted dataset forms the basis for fair comparisons among various recognition frameworks. Building upon the WiDar3.0 dataset, WiHF \cite{li2020wihf} derives a domain-independent motion change pattern for arm gestures, providing unique features for cross-domain recognition. WiSGP \cite{liu2023generalizing} employs a data augmentation-based approach to augment data and domain information, achieving a Wi-Fi gesture recognition system with domain-generalization capabilities. WiSR \cite{liu2023wisr} leverages the differences between subcarriers to accomplish domain-generalization in Wi-Fi gesture recognition. WiGRUNT \cite{gu2022wigrunt}, PAC-CSI \cite{su2023real} and \cite{gu2023attention}, on the other hand, focus on attention mechanisms designed to automatically uncover critical information for gesture recognition. Wi-learner \cite{feng2022wi} utilizes autoencoders to enable small-sample cross-domain Wi-Fi gesture recognition. However, these approaches have not taken into account the more practical scenario of open-set gesture recognition, where considering all test samples as known poses significant potential risks.

\subsection{Identify Unknows in Wi-Fi}
In Wi-Fi sensing research, apart from gesture recognition, some efforts have touched upon the challenge of distinguishing unknown classes. However, these efforts are often mentioned as a part of a larger study and lack in-depth analysis of this specific challenge. The identification of unknown classes can be achieved by utilizing the scores provided by classifiers \cite{bendale2015towards, mendes2017nearest}. For instance, setting a threshold on classification scores from traditional machine learning models \cite{scheirer2012toward} or deep softmax learners \cite{zhou2021learning,bendale2016towards,hendrycks2016baseline} can be used to determine whether an input sample belongs to an unknown class. Nevertheless, this approach is most suitable for identifying unknown samples situated near the classifier's decision boundary. It can often lead to misclassification when dealing with samples far from the classification boundary \cite{sun2020conditional}.

Existing studies on Wi-Fi-based unknown class recognition have adopted another approach, known as the prototype-based method \cite{yang2020convolutional,chen2020learning,chen2021adversarial,huang2022class}. In the works of Wione \cite{gu2021wione} and CAUTION \cite{wang2022caution}, a conventional softmax classifier is used for user identity classification. After model training, they summarize a prototype for each class. When a sample is equidistant from prototypes of two or more classes, it is classified as an unknown user. However, this approach fails when an unknown sample is close to one known class but far from others. Tan et al. \cite{tan2020enabling} and Multitrack \cite{tan2019multitrack} directly learn a prototype for each activity class. Activity recognition is achieved by comparing the similarity of input samples with prototypes of all classes. When the lowest similarity with all prototypes falls below a certain threshold, the sample is classified as unknown. This method does not consider compressing the feature space of known classes to reduce open space risk. Moreover, these approaches assume that each class can be represented by a single prototype. While this assumption is applicable in visual domains, the unique uncertainty in Wi-Fi sensing signals results in highly irregular data distributions. Consequently, prototypes obtained through traditional methods are inadequate representations of corresponding classes. Our proposed method, WiOpen, acknowledges the impact of uncertainty, conducts in-depth analysis, and designs targeted solutions, thus achieving Wi-Fi-based OSGR.

\section{Preliminaries}

In this section, we first introduce the concepts of open-set recognition and open space risk. We then discuss the uncertainty in Wi-Fi based open-set gesture recognition. Finally, we propose the motivation of WiOpen.

\subsection{Open-Set Recognition and Open Space Risk}
Open-set recognition (OSR) pertains to situations where new, previously unseen classes emerge during testing. In such cases, a classifier should not only accurately classify known classes but also effectively reject unknown ones. Considering a training set $\mathbf{D_L}=\{(x_1,y_1),...(x_n,y_n)\}$ of $n$ labeled gesture samples, where $x_i$ represents each sample and $y_i\in \{1...Y\}$ denotes the label of $x_i$, and a testing dataset $\mathbf{D_T}=\{(t_1),...(t_u)\}$ where the label of $t_i$ belongs to $\{1...Y+Q\}$, with $Q$ representing the number of unknown classes typical in real-world scenarios. The open space, denoted as $\mathcal{O}$, encompasses regions far from the known classes. And the degree of openness, which quantifies the open space in the OSR, can be described as $\mathcal{P}$ \cite{geng2020recent}:
\begin{equation}
    \mathcal{P}=1-\sqrt{\frac{2\times Y}{2*Y+Q} } 
\label{eqr:op}
\end{equation}

Inevitably, designating any sample in the open space as a known class introduces risks, known as open space risk ($\mathcal{R}_o$). Qualitatively, $\mathcal{R}_o$ can be described as the relative measure of the open space $\mathcal{O}$ in comparison to the overall measurement space $\mathcal{M}$ \cite{scheirer2012toward}:
\begin{equation}
    \mathcal{R}_\mathcal{O}=\frac{\int_\mathcal{O} f(x)dx}{\int_{\mathcal{M}} f(x)dx} 
\label{equ:ro}
\end{equation}
Here, $f(x)$ represents the measurable recognition function, where $f(x) = 1$ denotes that a certain class within the known classes has been recognized,  otherwise $f(x) = 0$. In other words, the more samples from the open space are classified as known classes, the higher $\mathcal{R}_\mathcal{O}$ becomes.

In the context of OSR and considering the concepts of open space risk and openness, the fundamental requirement for addressing the OSR problem is to determine a recognition function, denoted as $f(x)$, that minimizes the following open-set risk:
\begin{equation}
    arg \min_{f} \left \{ \mathcal{R}_\mathcal{O}(f,\mathbf{D_U}) + \lambda _r \mathcal{R}_\mathcal{E}(f,\mathbf{D_L})\right \} 
    \label{equ:osr}
\end{equation}
Here, $\lambda _r$ serves as a regularization constant, and $\mathcal{R}_\mathcal{O}$ and $\mathcal{R}_\mathcal{E}$ represent the open space risk and the empirical risk (the risk associated with incorrectly assigning known samples), respectively. $\mathbf{D_L}$ denotes the set of known labeled training data, and $\mathbf{D_U}$ represents the potentially unknown data.

\subsection{When Wi-Fi Based Gesture Recognition Meets Open-Set Challenge}

Current Wi-Fi based sensing solutions predominantly rely on Channel State Information (CSI) \cite{huang2021phaseanti,liu2021adaptive,yan2019wiact}. CSI characterizes the signal attenuation that occurs as signals propagate through a given medium. This attenuation can be expressed through the following equation:
\begin{equation}
	\label{equ:CH}
	Y= H X+ \mathcal{N}
	\end{equation}
Where $Y$ and $X$ are the received and transmitted signal vectors, respectively. $\mathcal{N}$ is additive white Gaussian noise, and $H$ stands for the channel matrix representing the CSI.

CSI can be delineated as the combination of two main components: static CSI and dynamic CSI. Static CSI is influenced by the surrounding environment and the presence of a Line-of-Sight (LoS) between the transceivers. Dynamic CSI, on the other hand, is primarily determined by the reflection path from moving objects:
\begin{equation}
\label{equ:CFRSUM}
H(r,t)=H_s(r,t)+H_d(r,t)
\end{equation}
Where $r$ and $t$ represent the signal frequency and the timestamp, respectively. The dynamic CSI can be further elaborated as:
\begin{equation}
\label{equ:CFR}
H_d(r,t)=\sum_{k\in \mathbf{D}} h_k(r,t) e^{-j2\pi \frac{ d_k(t)} {\lambda_k}}
\end{equation}
Here, $h_k(r,t)$, $d_k(t)$, and $\lambda_k$ represent the attenuation, the path length of the dynamic path, and the wavelength associated with the $k^{th}$ path, respectively. The set $\mathbf{D}$ encompasses dynamic paths. Notably, gestures induce changes in length of the dynamic paths, subsequently altering the overall CSI sequence. Traditional Wi-Fi-based gesture recognition techniques aim to extract gesture-related information from the overall CSI and subsequently interpret and recognize the original gestures.


In contrast to close-set methods, Wi-Fi-based OSGR aims to accomplish the dual task of recognizing known gestures accurately while effectively identifying unknown gestures. Equation \ref{equ:osr} highlights that the challenge in solving the Wi-Fi OSGR problem involves the simultaneous minimization of empirical classification risk for labeled known data and open-space risk for potential unknown data. Open-space risk, as quantified by Equation \ref{equ:ro}, and empirical risk can be expressed as follows:
\begin{equation}
    \mathcal{R}_\mathcal{E}=\frac{1}{N} \sum_{i=1}^{N} \mathcal{L}(f(x_i),y_i)
    \label{equ:er}
\end{equation}
Here, $N$ signifies the number of training samples, while $x_i$ and $y_i$ denote the $i$-th sample and its corresponding label. The loss function, $\mathcal{L}$, gauges the likelihood of the prediction $f(x_i)$ with respect to the true label $y_i$. 

It is important to note that both $\mathcal{R}_\mathcal{O}$ and $\mathcal{R}_\mathcal{E}$ are inherently linked to the misclassification tendencies of the gesture recognition model, $f(x)$. Traditional recognition models typically transform input data from the data space to a feature space and subsequently establish a decision boundary within this feature space to differentiate between distinct classes. Consequently, misclassification may be attributed to three key factors: 1. Data confusion in the data space, 2. Feature confusion in the feature space, 3. Overlapping decision boundaries. The extent of data and feature confusion essentially determines the upper bounds of decision boundary learning. Furthermore, $\mathcal{R}_\mathcal{O}$ is also influenced by the establishment of decision boundaries used for identifying unknowns. $\mathcal{R}_\mathcal{O}$ performs best when this boundary only appropriately encompasses all samples of known classes. However, unfortunately, due to the impact of the unique uncertainty inherent in Wi-Fi sensing, both reducing data confusion and determining suitable decision boundaries face significant challenges.



\begin{figure*}[ht]
\centering
\includegraphics[width=2\columnwidth]{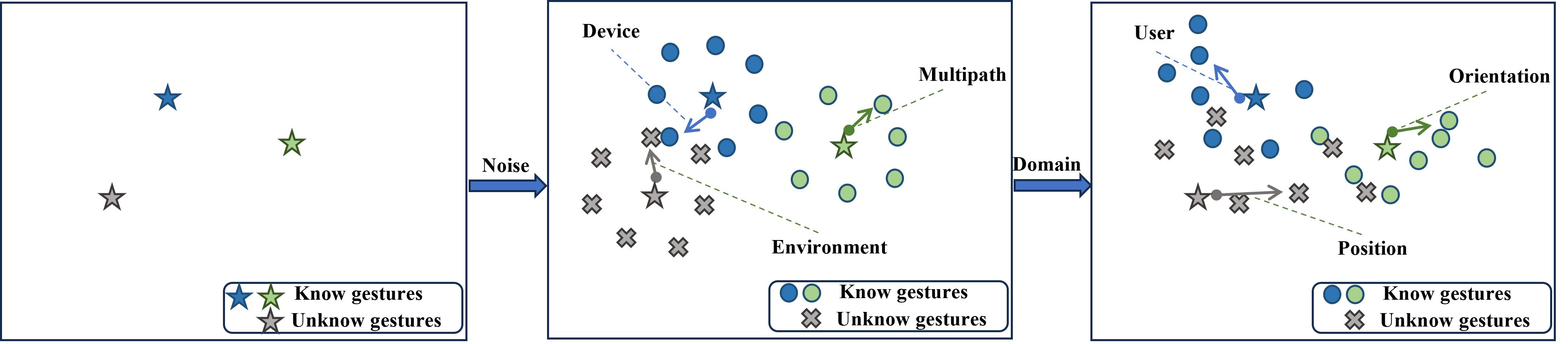}
\caption{The multipath effect, variations in the environment, changes in the position and orientation of the performer, and interference all contribute to an increased randomness in the distribution of Wi-Fi CSI data.}
\label{fig:unc}
\end{figure*}

\begin{figure}
    \centering
    \includegraphics[width=0.95\linewidth]{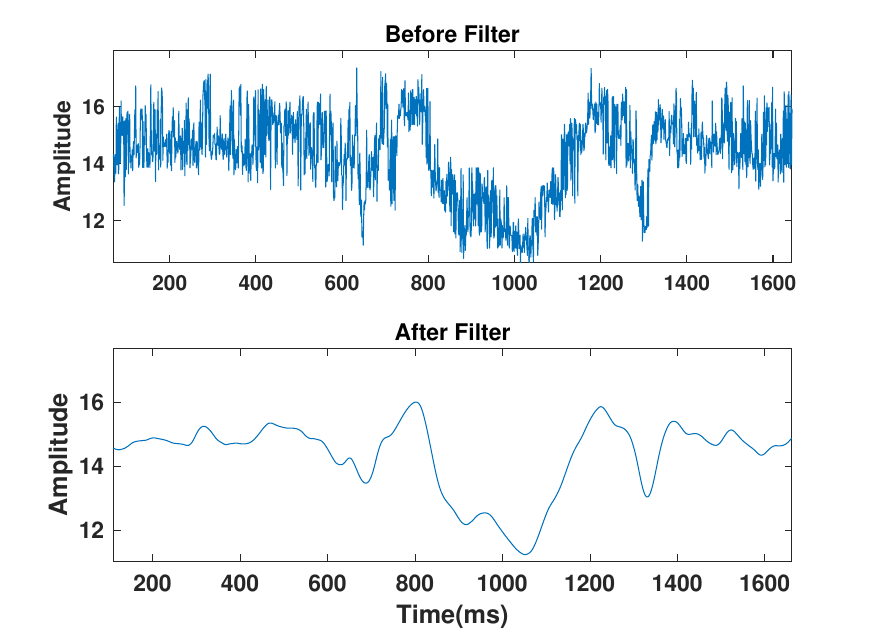}
    \caption{Examples of samples with different noise uncertainty, where $\Psi_{n}$ before filtering is 1.92, and $\Psi_{n}$ after filtering is 0.96.}
    \label{fig:filter}
\end{figure}

We define uncertainty in Wi-Fi sensing as \textbf{the intrinsic variability, randomness, or ambiguity present in the available sensing data}. In a nutshell, uncertainty introduces a certain degree of bias to sensory data, resulting in a spatial displacement of samples in the data space, ultimately leading to irregular and discrete distributions of collected CSI samples. We posit that uncertainty in Wi-Fi sensing arises from two factors: noise and domains (such as distinct users or different gesture execution locations), and it can be expressed as:
\begin{equation}
    \Psi=\Psi_{n}+\Psi_{d}
    \label{equ:unc}
\end{equation}
where $\Psi$, $\Psi_{n}$ and $\Psi_{d}$ represents the overall uncertainty, the uncertainty caused by noise and the uncertainty caused by domains.

A portion of the noise present in CSI arises from multipath effects. From equations \ref{equ:CFRSUM} and \ref{equ:CFR}, the received CSI is the summation of signals from multiple propagation paths, which makes the target path signal susceptible to interference from unrelated environmental signals \cite{zhang2023wital}. Additionally, noise is also introduced by Wi-Fi devices themselves, contributing to $\Psi_{n}$:
\begin{equation}
    \Psi_{n} = \Psi_{n}^e + \Psi_{n}^d
\end{equation}
Where $\Psi_{n}^e$ and  $\Psi_{n}^d$ represent the noise uncertainty caused by environment and devices. Noise introduces random biases to the collected CSI samples, causing an expansion in the distribution range of these samples. This results in larger intra-class differences and smaller inter-class differences, ultimately making it easier for confusion to occur between classes, as shown in Figure \ref{fig:unc}.

Domain-induced uncertainty similarly introduces biases to the collected CSI samples. However, unlike noise uncertainty, the biases introduced by domain uncertainty are directional, closely related to the principles of Wi-Fi sensing. When there is no significant change in static paths, CSI variations are primarily driven by dynamic path changes caused by activities. According to the Fresnel zone theory\cite{wang2016human,niu2018fresnel,zhang2019towards}, even if the user's gestures remain unchanged, variations in user positions and orientations concerning the transmitting and receiving antennas lead to corresponding changes in the dynamic path variations. Wi-Fi sensing is highly sensitive to domain changes, and as CSI data represents the superimposition of signals from multiple paths, it possesses a lower spatial resolution compared to sensory data in domains such as vision. Domain changes result in substantial data distribution shifts towards the corresponding directions, as illustrated in Figure \ref{fig:unc}. This type of uncertainty is referred to as $\Psi_{d}$:
\begin{equation}
    \Psi_{d} =  \Psi_{d}^u + \Psi_{d}^l +...\Psi_{d}^{nd}
\end{equation}
Where $\Psi_{d}^u$, $\Psi_{d}^d$ and $\Psi_{d}^{nd}$ represent user, location and $nd$ th domain. Such uncertainty is common in Wi-Fi based sensing, leading to a widely scattered and irregular distribution of CSI data \cite{zhang2021widar3, li2020wihf}.


It can be observed that addressing the challenges of Wi-Fi-based OSGR requires eliminating the effects of uncertainty, and before eliminating uncertainty, it is advisable to quantify it. Noise adds random noise to CSI data, and the data distribution of this random noise should follow a normal distribution with an unknown variance. Therefore, quantifying noise uncertainty can be transformed into assessing the proportion of data in CSI that conforms to this normal distribution and the magnitude of the variance of this normal distribution. On the other hand, domain uncertainty results in directional biases in the data. When domain uncertainty is higher, the distance of an individual CSI sample to other samples of the same class increases, while the distance to samples of other classes decreases. Thus, it can be transformed into assessing the distances between all samples and other samples. Therefore, we propose the following formula to quantify uncertainty.
\begin{equation}
\begin{aligned}
    \Psi&=\Psi_{n}+\Psi_{d}\\
           &=\frac{1}{N*G}(\sum_{n=1}^{N} (\sum_{k=1}^{G} (\alpha _{nk}- \bar{\alpha _n})^2  + \sum_{k=1}^{G} \sigma _{nk}))\\
           &+\frac{1}{N}(\sum_{n=1}^{N}\sum_{i=1}^{N}d(x_n,x_i)-\sum_{n=1}^{N}\sum_{j=1}^{N}d(x_n,x_j))
    \label{equ:uncerqu}
\end{aligned}
\end{equation}
Here, $i \ne n$ and $j \ne n$, $y_i \ne y_n$ and $y_j = y_n$.

The two terms in the formula above serve as measures of noise uncertainty and domain uncertainty. For the first term, we fit a CSI sample, denoted as $x_n$, into a Gaussian Mixture Model (GMM) with $G$ components, leading to the following probability distribution:
\begin{equation}
    p(x_{nt})=\sum_{k=1}^{G} \alpha _{nk}\eta (x_{k},\mu _{nk},\sigma _{nk})
\end{equation}
Where $\alpha _{nk}$ represents the probability that the observed data belongs to the $k$th component, $\mu _{nk}$ and $\sigma _{nk}$ represent the expectation and variance of the $k$th component, respectively. Here, $\eta$ is the Gaussian probability density function and $N$ is the number of samples. \(x_{nk}\) represents the reading at the $k$th timestamp of the $n$th sample. When $x_n$ exhibits larger noise, it indicates stronger random noise. This results in one of $K$ Gaussian components (representing random noise) having a higher $\alpha _{nk}$. Moreover, greater noise leads to a more random data distribution, resulting in a larger $\sigma _{nk}$. Consequently, we design the first term to quantify  $\Psi_{n}$, and the effect is shown in Figure \ref{fig:filter}.

\begin{figure}
	\centering
		\subfloat[]{\label{dunl}
		\begin{minipage}{0.46\linewidth}
			\centering
			\includegraphics[width=1\textwidth]{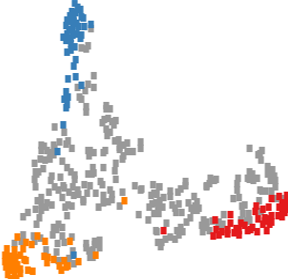}
		\end{minipage}
	}	
	\quad
	\subfloat[]{\label{dunm}
		\begin{minipage}{0.46\linewidth}
			\centering
			\includegraphics[width=1\textwidth]{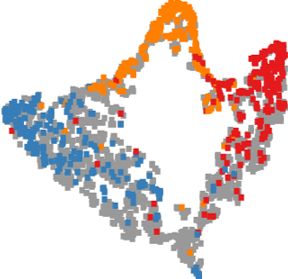}
		\end{minipage}
	}	
	\quad
	\caption{Domain uncertainty.(a) 0.0901; (b) 0.2659. Colored points represent samples from known classes, while gray points represent samples from unknown classes.}
	\label{fig:dun}
\end{figure}

The second term in Equation \ref{equ:uncerqu} is used to measure $\Psi_{d}$. In this term, $y_n$ represents the label of $x_n$, and $d$ denotes the distance metric function. The objective of this term is to gauge domain-related uncertainty by considering neighborhood relationships. In essence, when there is a lower probability that closer neighboring samples belong to the same class, uncertainty increases. Conversely, the higher the probability that a sample's neighbors originate from other classes, the greater $\Psi_{d}$ becomes. We selected some data from the WiDar3.0 \cite{zhang2021widar3} dataset and presented their data distribution as shown in Figure \ref{fig:dun}. Figure \ref{dunl} includes data from different users, while Figure \ref{dunm} includes data from different users, directions, and positions. It can be observed that domain factors do increase the irregularity in data distribution. Our uncertainty measurement approach can also quantify domain uncertainty.

\subsection{Motivation of WiOpen}
Based on the previous analysis, effectively addressing the open-set challenge in Wi-Fi-based gesture recognition hinges on mitigating the impact of uncertainty in data/feature space. To mitigate uncertainty, we propose a method that involves designing strategies in both the data preprocessing and feature learning stages, each serving a distinct purpose. In the data preprocessing stage, WiOpen focuses on the reduction of noise interference. This step is essential in eliminating unwanted data variations introduced by environmental factors, device noise, and multipath effects. In the feature learning stage, our approach utilizes the uncertainty quantification method-inspired OSGR network to facilitate dynamic-related feature extraction and sample clustering in the feature space. These features aim to create more compact representations, which in turn significantly lower uncertainty, thus enhancing the final decision-making process. Specifically, according to the uncertainty quantification method, the OSGR network assesses the distances between each sample and all other samples, effectively bringing closer the distances between samples of the same class and pushing away those from other classes during training.

In the decision-making phase, it is essential to learn decision boundaries to effectively bound open space risk and empirical risk under the impact of uncertainty. Deep learning techniques, have demonstrated promise in various Wi-Fi-based gesture recognition tasks. Traditional softmax classifiers \cite{li2020wihf, zhang2021widar3} are efficient at measuring empirical risk, but they have limitations. The decision boundary they establish partitions the entire feature space into regions, equal to the number of known classes. Furthermore, softmax scores normalize measures that quantify the ratio of the distance between a sample and a class to the sum of distances between the sample and all other classes. Thus softmax classifiers may not effectively bound open space risk. Even for an unknown sample, if its distance to a known class is significantly closer than to other known classes, the softmax scores would assign a high probability to it belonging to the closest known class. This results in a classifier that only rejects sample near the decision boundary, causing regions of acceptance to be infinitely wide, as illustrated in Figure \ref{fig:intro}.

Furthermore, while the proposed approach can eliminate a portion of uncertainty, the final samples feature distribution still exhibits irregularity. Therefore, the prototype-based approaches continues to face challenges in obtaining optimal decision boundaries. To address this issue, building upon the uncertainty-inspired OSGR network, we introduce decision functions constructed based on semantic relationships between samples and their neighbors. Specifically, distances between samples are used to define a classification distribution based on class labels. Neighbor relationship mining helps overcome the limitation of learning a single protype for each class, making it more suitable for Wi-Fi sensing data with significant irregularity. Furthermore, employing uncertainty measurement (distance in WiOpen) directly as the decision boundaries optimization parameter makes the network more conducive to reducing uncertainty, achieving end-to-end uncertainty reduction, and effective decision boundary planning \cite{scheirer2014probability}. 

\section{Method}

\begin{figure}
\centering
\includegraphics[width=0.95\columnwidth]{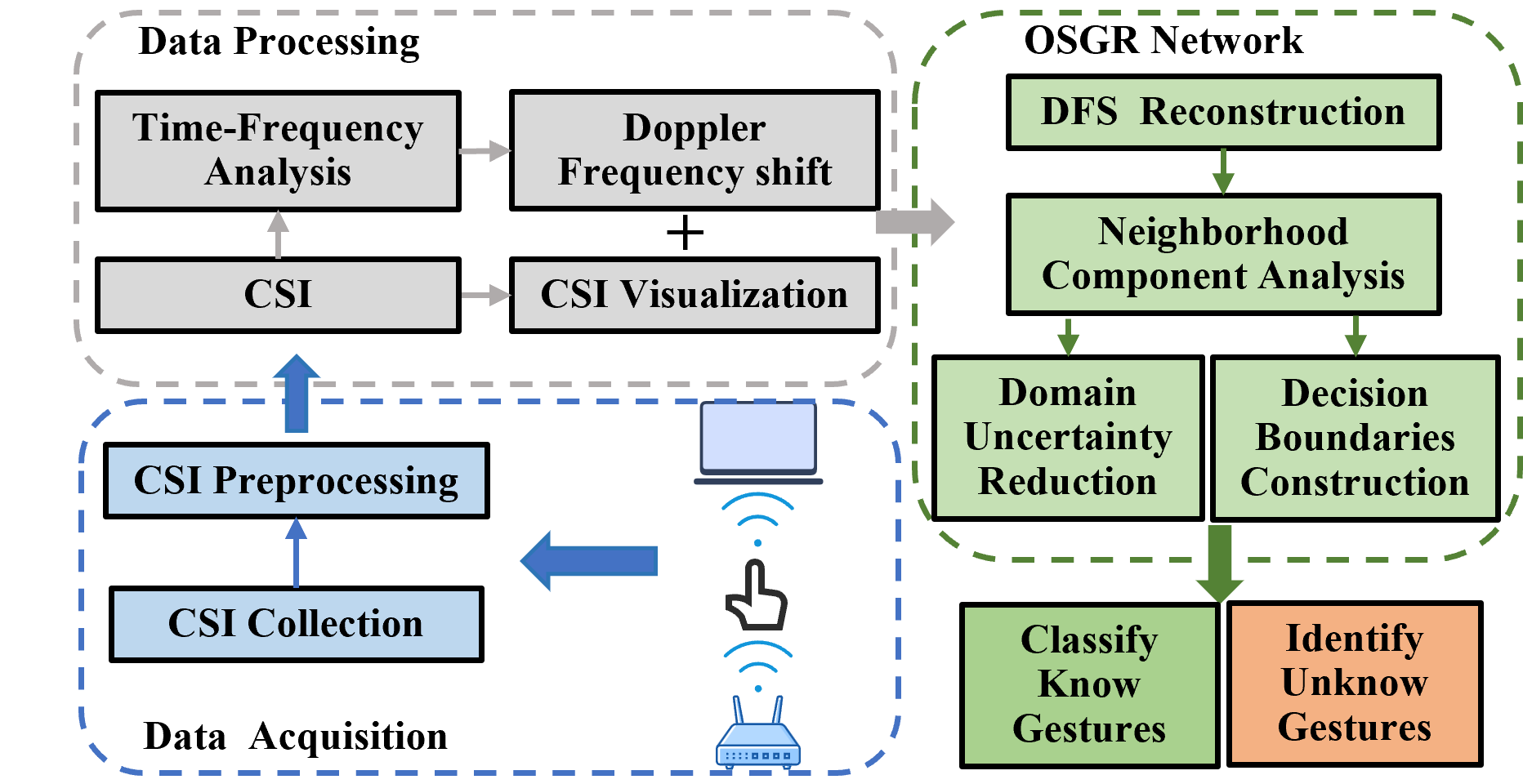}
\caption{WiOpen System.}
\label{fig:met}
\end{figure}

\subsection{Overview of WiOpen}

As shown in Figure \ref{fig:met}, WiOpen is composed of three main components: data acquisition, data processing, and the OSGR network. The data acquisition component is responsible for obtaining data from Wi-Fi sensing devices and performing data preprocessing to eliminate some of the noise uncertainty. The data preprocessing component involves extracting DFS from CSI and visualizing it along with CSI amplitude and phase as input to the OSGR network. The OSGR network utilizes the reconstruction from CSI to DFS to eliminate influence of static paths, and learns the semantic neighborhood structure of samples to eliminate domain uncertainty and construct decision boundaries for open-set recognition.

\subsection{Data Acquisition}
The CSI is collected from wireless network cards capable of capturing CSI. As demonstrated in the previous section, the gesture can be portrayed by the change CSI. However, with commodity Wi-Fi devices, there's a challenge due to unsynchronized transmitters and receivers, leading to a time-varying random noise $e^{-j\theta_{\text{n}}}$:
\begin{equation}
\begin{aligned}
\label{equ:off}
H(f,t)&= e^{-j\theta_{n}}(H_{s}(f,t)+ H_{d}(f,t))\\
&=\ e^{-j\theta_{n}}(H_{s}(f,t)+ A(f,t) e^{-j2\pi \frac{d(t) }{\lambda } })
\end{aligned}
\end{equation}
where $A(f,t)$, $e^{-j2\pi \frac{d(t) }{\lambda }}$ and $d(t)$ denote the complex attenuation, phase shift and path length of dynamic components, respectively. This random phase noise, $e^{-j\theta_{\text{n}}}$, hinders the direct use of CSI phase information and increase noise uncertainty.

Therefore, we need to eliminate $e^{-j\theta_{n}}$. Fortunately, for commodity Wi-Fi cards, this noise remains constant across different antennas on the same Wi-Fi Network Interface Card (NIC) because they share the same RF oscillator. This noise can be eliminated using the CSI-ratio model \cite{wu2020fingerdraw}:
\begin{equation}
	\begin{aligned}
		\label{equ:csiratio}
		H_q(f,t) &= \frac{H_{1}(f,t)}{H_{2}(f,t)} \\
		&=\frac{e^{-j\theta _{n}}(H_{s,1}+A_{1}e^{-j2\pi \frac{d_{1}(t)}{\lambda } })}{e^{-j\theta _{n}}(H_{s,2}+A_{1}e^{-j2\pi \frac{d_{2}(t)}{\lambda } })}\\
		&=\frac{A_{1}e^{-j2\pi \frac{d_{1}(t)}{\lambda }} +H_{s,1}}{A_{2}e^{-j2\pi \frac{d_{1}(t) + \bigtriangleup d }{\lambda } } +H_{s,2}}
	\end{aligned}
\end{equation}
where $H_{1}(f,t)$ and $H_{2}(f,t)$ are the CSI of two receiving antennas. When two antennas are close to each other, $\bigtriangleup d$ can be regarded as a constant. According to Mobius transformation \cite{zeng2019farsense}, equation \ref{equ:csiratio} represents transformations such as scaling and rotation of the phase shift $e^{-j2\pi \frac{d_{1}(t)}{\lambda }}$ of antenna 1 in the complex plane, and these transformations will not affect the changing trend of the CSI. Following the CSI ratio-processing, we further mitigate noise uncertainty by subjecting the data to a low-pass filter for the removal of environmental noise. Subsequently, the processed data undergoes further refinement in the data processing section.

\subsection{Data Processing}

\begin{figure}
    \centering
    \includegraphics[width=0.6\linewidth]{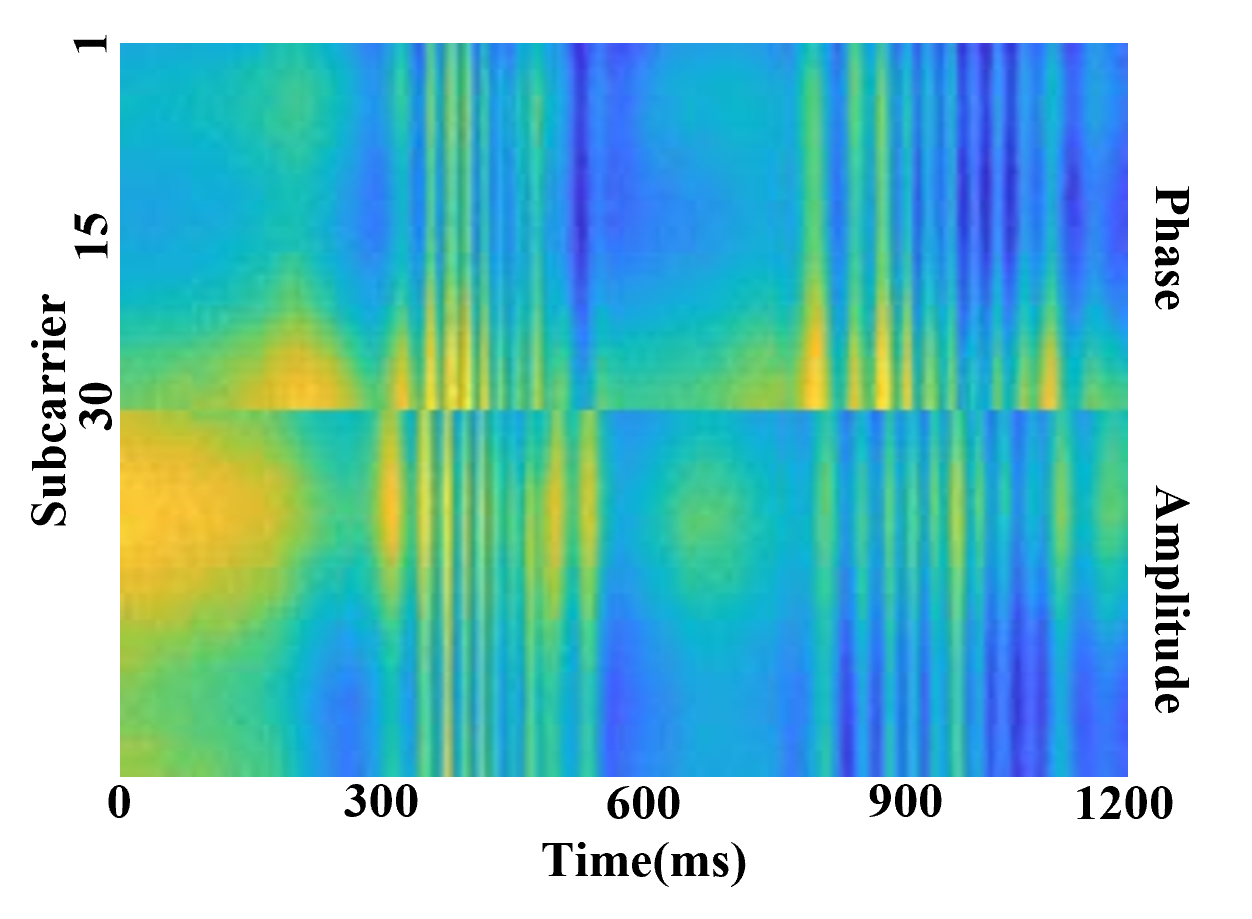}
    \caption{The amplitude and phase visualization for a sample of "push \& pull" gesture.}
    \label{fig:amp}
\end{figure}

The data processing section involves visualizing the CSI and extracting DFS, which are fundamental for providing high-quality inputs to the OSGR network. Specifically, To ensure that the OSGR network receives normalized and information-rich inputs, we adopt the CSI visualization technique detailed in \cite{gu2022wigrunt}. This method visualizes both the amplitude and phase of CSI separately and then integrates them into a single image. The advantage of this approach is that it consolidates the various dimensions of CSI information into a two-dimensional matrix, which is suitable for processing using a Convolutional Neural Network (CNN)-based network. The visualization results for CSI amplitude and phase are depicted in Figure \ref{fig:amp}.

\begin{figure}
	\centering
		\subfloat[]{\label{dfsr}
		\begin{minipage}{0.46\linewidth}
			\centering
			\includegraphics[width=1\textwidth]{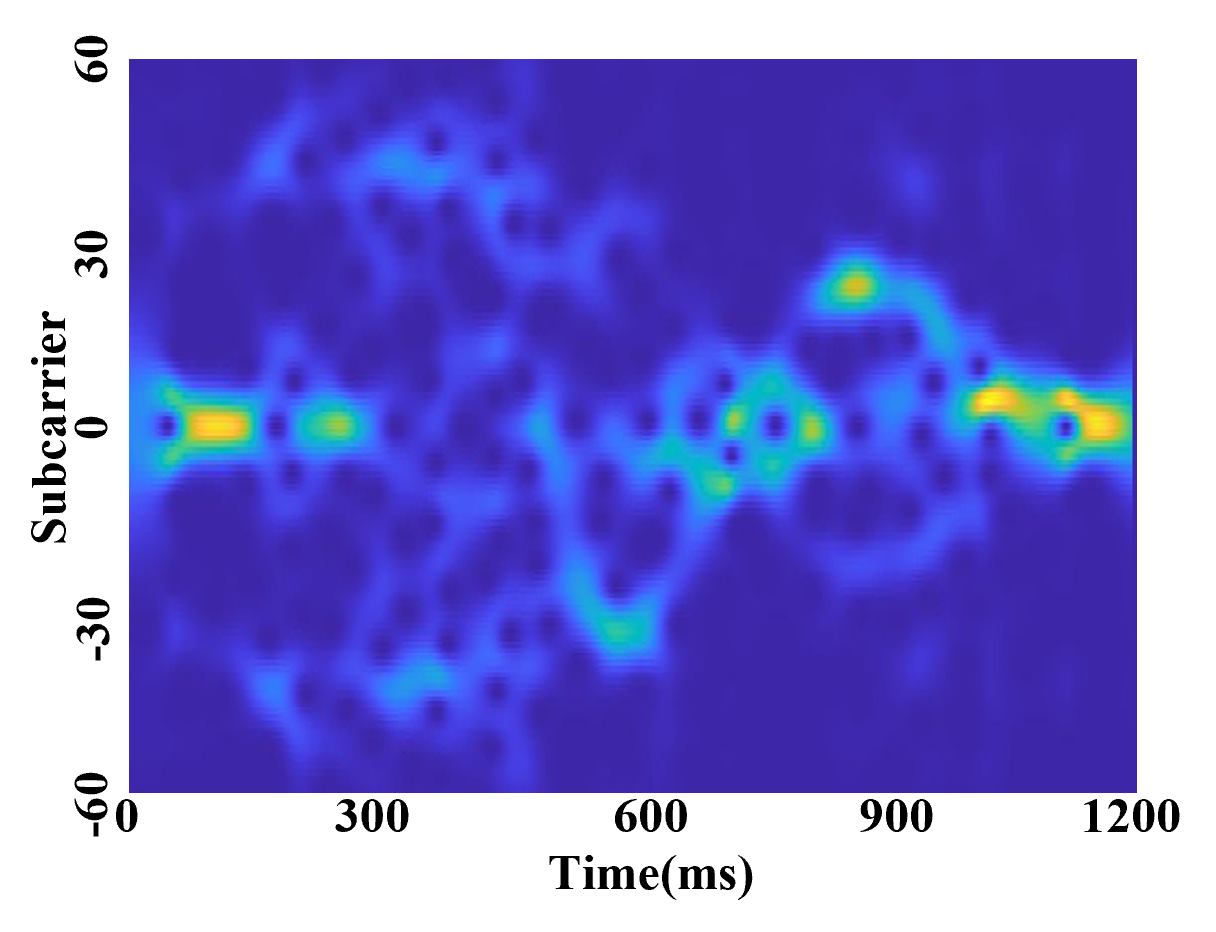}
		\end{minipage}
	}	
	\quad
	\subfloat[]{\label{dfs}
		\begin{minipage}{0.46\linewidth}
			\centering
			\includegraphics[width=1\textwidth]{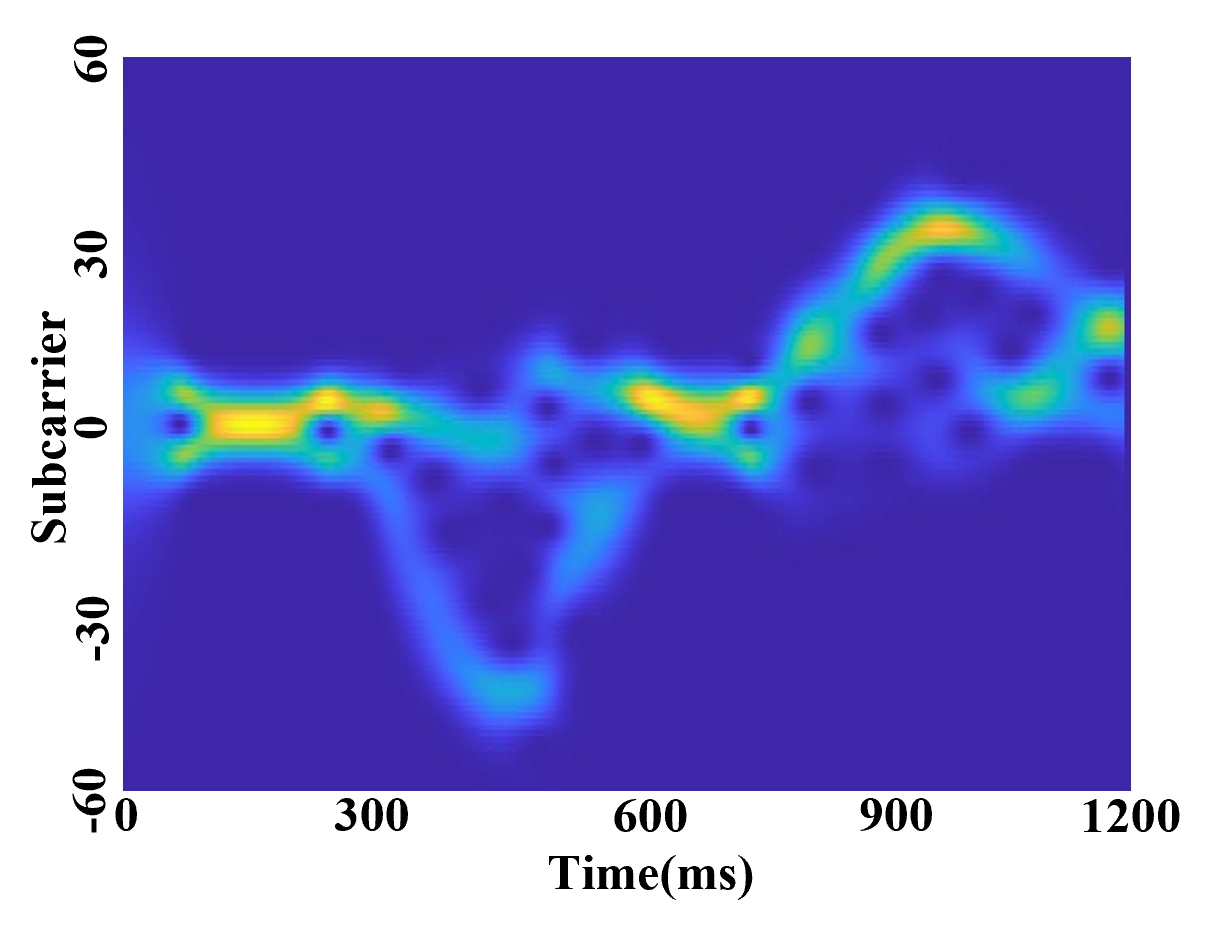}
		\end{minipage}
	}	
	\quad
	\caption{The DFS visualization for a sample of "push \& pull" gesture.(a) Previous method\cite{zhang2021widar3}; (b) Our method. }
	\label{fig:dfs}
\end{figure}

In contrast to previous methods, WiOpen derives DFS from data processed with CSI ratio, as opposed to CSI conjugate multiplication \cite{zhang2021widar3}. The CSI ratio method is advantageous because it represents transformations such as rotation and scaling of CSI in the complex plane, thus avoiding some negative effects introduced by $H_{2}(f,t)$. To further mitigate cumulative error effects caused by $\bigtriangleup d$, we apply an antenna selection coefficient $s_a$ to select $H_{1}(f,t)$ and $H_{2}(f,t)$. This coefficient is calculated as follows:
\begin{equation}
    s_a = \frac{1}{C}\sum_{c=1}^{C}\frac{var(|H_a(f_c,t))|}{mean(|H_a(f_c,t)|)}   
\end{equation}
where $var$ and $mean$ denote the variance and mean value of amplitude readings for the $a$th antenna of the $c$th subcarrier. We select the antennas with the highest and lowest $s_a$ values as $H_{1}(f,t)$ and $H_{2}(f,t)$, respectively. The rationale behind this selection is that CSI with larger variances is generally more sensitive to motion, while CSI with higher amplitude typically contains a larger static path component, making $H_{1(f,t)}$ less affected by $\bigtriangleup d$.

After obtaining denoised CSI after data acquisition, and to further reduce the impact of the static path (without introducing DFS), we apply a high-pass filter. Finally, we utilize Fast Fourier Transform (FFT) to obtain DFS. A comparison between the method used to extract DFS in previous approaches and the one adopted by WiOpen is illustrated in Figure \ref{fig:dfs}. Notably, the latter produces higher-quality DFS.

\subsection{Open-Set Gesture Recognition Network}

\begin{figure*}[ht]
    \centering
    \includegraphics[width=1\linewidth]{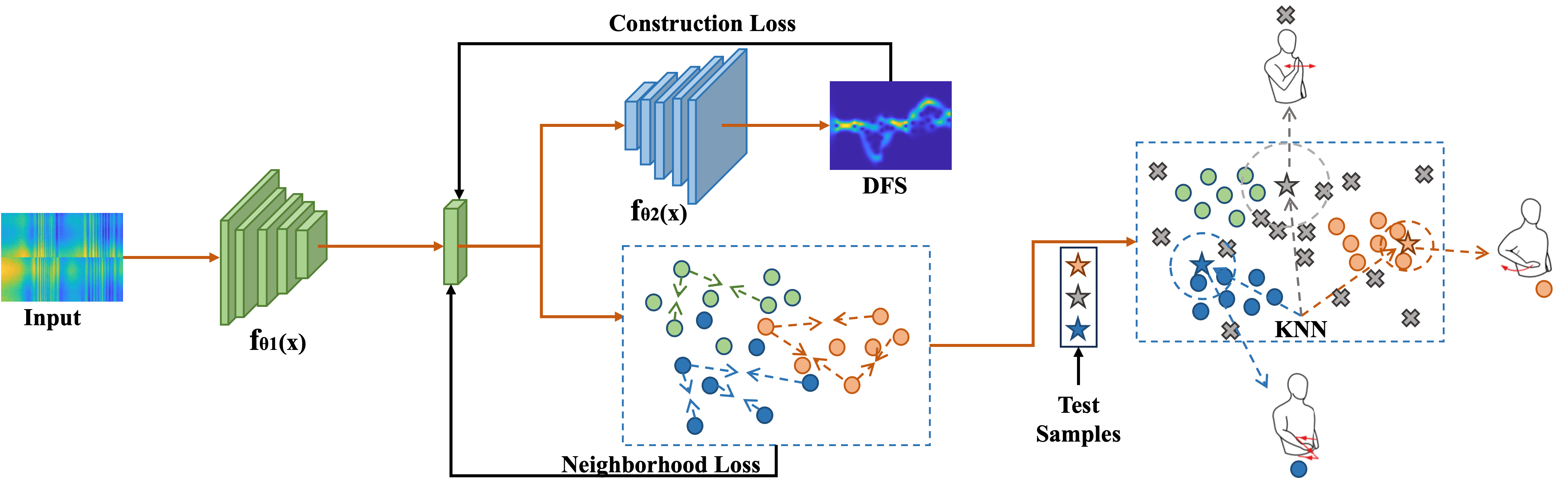}
    \caption{Framework of the proposed open-set gesture recognition network.}
    \label{fig:osgr}
\end{figure*}

The proposed OSGR network, depicted in Figure \ref{fig:osgr}, is structured into two branches, each with specific objectives. The first branch is dedicated to constructing DFS outputs from the original inputs, while the other is focused on learning valuable class-related knowledge. Throughout the training process, these branches are guided by the neighborhood loss and construction loss, respectively, to facilitate effective knowledge acquisition. During testing, a K-Nearest Neighbors (KNN)-based decision method is employed to classify known samples and reject unknown ones.

Even after data preprocessing, the CSI retains influences from static path interference and domain uncertainty. To address the former, some existing approaches \cite{li2020wihf, zhang2021widar3, feng2022wi} advocate using DFS or its derivatives as input for the network, as DFS predominantly captures dynamic path characteristics and effectively mitigates interference from static paths. However, relying solely on DFS as input may result in the loss of valuable information. Therefore, our approach retains both amplitude and phase as input but incorporates a construction loss to guide the network in leveraging dynamic path-related information and mitigating interference from static paths to the fullest extent. As illustrated in Figure \ref{fig:osgr}, $f_{\theta 1}(x)$ represents the backbone network responsible for feature learning. It extracts low-dimensional features, denoted as $v_n$, from the sample. These features are then transformed into DFS corresponding to the sample through the first branch $f_{\theta 2}(x)$. During the process of constructing DFS, a bottleneck structure exists between $f_{\theta 1}(x)$ and $f_{\theta 2}(x)$. This structure ensures that the learned features $v_n$ contain as much information related to dynamic paths as possible to achieve high-quality DFS construction. Therefore, training this branch using the construction loss serves the purpose of eliminating influence associated with static paths. The construction loss is defined as:

\begin{equation}
    \mathcal{L}_c = MSE(x_n^{dfs},f_{\theta 2}(f_{\theta 2}(x_n)))
\end{equation}
Where $MSE$ represents Mean Squared Error loss, $x_n$ is the input sample, and $x_n^{dfs}$ is the DFS corresponding to $x_n$.

For addressing domain uncertainty, drawing inspiration from the uncertainty quantification method outlined in Equation \ref{equ:uncerqu}, we have devised a scheme based on neighbor component analysis (NCA) \cite{wu2018improving}. In this approach, NCA computes the distance of each training sample to all other samples within the embedding space. Subsequently, guided by the distance metric and class labels, it minimizes the distances between samples of the same class while pushing samples from other classes further apart, thus effectively reducing domain uncertainty. The second branch is specifically dedicated to domain uncertainty reduction. Considering a training sample $x_n \in [x_1...x_N]$ along with its corresponding label $y_n$ and its feature vector $v_n = f_{\theta 1}(x_n)$, we employ similarity as the distance metric. The similarity between $x_n$ and another sample $x_j \in [x_1...x_N]$ is defined as the cosine similarity:
\begin{equation}
    s_{nj}=\frac{v_n^T}{||v_n||\space ||v_j||} = v_n^T v_j
\end{equation}
The probability that sample $x_n$ selects $x_j$ as its neighbor is:
\begin{equation}
    p_{nj} = \frac{exp(s_{nj}/\gamma)}{ {\textstyle \sum_{k\ne n}^{} exp(s_{nk}/\gamma)}}, p_{nn}=0
\end{equation}
$\gamma$ is the parameter to control the scale of the neighborhood. 

During the training process, for each sample $x_n$, the OSGR network computes its probability of being classified correctly as follows:
\begin{equation}
    p_n=\sum_{j\in S_n}^{} p_{nj}
\end{equation}
$S_n$ represents the indices of all samples with labels identical to that of $x_n$. The loss for the second branch, referred to as neighbor loss, is defined as:
\begin{equation}
    \mathcal{L}_e = \frac{1}{N}\sum_{n=1}^{N}\mathcal{L}_{e}^{n}=-\frac{1}{N}\sum_{n=1}^{N}log(p_n)
\end{equation}
The gradient calculation of Loss $L_e$ with respect to the feature vectors $v_n$ is as follows:
\begin{equation}
    \frac{\partial \mathcal{L}_e^n}{\partial v_n}=\frac{1}{\gamma }(\sum_{k\in S_n}^{}p_{nk}v_k-\sum_{k\in S_n}^{}\tilde{p} _{nk}v_k)
\end{equation}
where $\tilde{p} _{nk} = p_{nk}/\sum_{j\in S_n}p_{nj}$ is the normalized distribution within the class $y_n$. The overall loss of the OSGR network is:
\begin{equation}
    \mathcal{L} = \mathcal{L}_e + \lambda \mathcal{L}_c
\end{equation}
Where $\lambda$ is the hyperparameter controlling the impact of the construction branch. Throughout the training process, $\mathcal{L}$ induces samples of the same class to converge in the feature space while simultaneously distancing them from samples of different classes. This process not only eradicates domain uncertainty but also enables each sample to harness knowledge associated with all other samples, facilitating the learning of class-related information. Consequently, this leads to the formation of a feature space characterized by semantic separability. Utilizing this semantic space allows for the classification of known classes and the discrimination of unknown classes.

For recognize know gestures and reject unknows, all feature vectors of training samples are stored in a feature database, referred to as $T$. During the testing phase, a query sample $x_t$ has its feature vector $v_t$ extracted using the pre-trained network $f_{\theta 1}(x_t)$. Subsequently, $v_t$ is used to query a set of $K$ nearest samples, referred to as $S_k$, from $T$ based on a similarity measure. The majority class within this set is then defined as the candidate label:
\begin{equation}
    y_{t}^{c}=max(p_c|p_c=\frac{sum(y_j = c)}{K})
\end{equation}

The final label for $x_t$ is determined based on $y_{t}^{c}$ and a threshold $t$:
\begin{equation}
    y_t = \left\{\begin{matrix}y_{t}^{c},\sum_{j\in S_t^c}{d(x_t,x_j)}<t
 \\y_u,\sum_{j\in S_t^c}{d(x_t,x_j)}\ge t

\end{matrix}\right.
\end{equation}
Here, $y_u$ signifies that $x_t$ belongs to an unknown class. The threshold $t$ is defined as:
\begin{equation}
    t=\xi * \frac{1}{B} \sum_{b=1}^{B} {max(\sum_{j\in  S_r}{d(x_r,x_j))}}
\end{equation}
Where $B$ represents the total number of training batches, $x_r$ refers to all the samples in batch $b$, and $\xi$ is a hyper-parameter used to control the threshold. In contrast to previous approaches, the proposed OSGR network formulates uncertainty measurement, feature learning objective, and classifier decision criteria based on inter-sample neighborhood relationships. This approach integrates the optimization objectives for uncertainty reduction, meaningful feature learning, and decision boundary definition in an end-to-end manner. Setting decision boundaries through neighborhood relationships overcomes the limitations of prototype-based schemes, allowing for dynamic and flexible determination of decision boundaries based on the distribution of surrounding samples. Furthermore, the proposed threshold determination approach facilitates adaptive threshold setting during training, thereby enhancing the flexibility of decision boundary determination.

\begin{table*}[ht]
	\centering   	
	\caption{\label{widar3} \upshape Description of Widar3 dataset}
	\begin{tabular}{cm{1cm}<{\centering}m{8cm}<{\centering}m{1.3cm}<{\centering}m{1.5cm}<{\centering}m{1.3cm}<{\centering}}
	    \toprule
		Environments & No. of Users & Gestures & No. of Locations &  No. of Orientations & No. of Samples\\
		\hline
		1st (Classroom
) & 9 & 1: Push Pull; 2: Sweep; 3: Clap; 4:Slide; 5: Draw-O(Horizontal); 6: Draw-Zigzag(Horizontal); 7: Draw-N(Horizontal); 8: Draw-Triangle(Horizontal); 9: Draw-Rectangle(Horizontal); & 5 & 5 & 10125\\
		\hline
		2nd (Hall) & 4 & 1: Push Pull; 2: Sweep; 3: Clap; 4:Slide; 5: Draw-O(Horizontal); 6: Draw-Zigzag(Horizontal); & 5 & 5 & 3000\\
		\hline
		3rd (Office) &  4 & 1: Push Pull; 2: Sweep; 3: Clap; 4:Slide; 5: Draw-O(Horizontal); 6: Draw-Zigzag(Horizontal); & 5 & 5 & 3000\\
		\bottomrule
	\end{tabular}
\end{table*}

\section{Experimental Analysis}

\subsection{Datasets}

In this section, we introduce two datasets, Widar3.0 and ARIL, which are used for experimental evaluation.

\textbf{Widar3.0:} The public dataset WiDar3.0 \cite{zhang2021widar3} contains $16125$ samples collected from 3 environments, and its detailed description is shown Table \ref{widar3}. To verify the performance of the WiOpen system, in section \ref{sec:ove}, we evaluate the know samples recognition and unknow samples rejection performance under different openness $\mathcal{P}$ with all the data from $1$st environment (10125 samples, 9 users  $\times$ 5 positions  $\times$ 5 orientations  $\times$ 9 gestures  $\times$ 5 instances). In section \ref{sec:cro}, we also use all samples to evaluate the in-domain and cross-domain performance of WiOpen same as other state-of-the-art researches \cite{liu2023generalizing,su2023real}.

\textbf{ARIL:} The ARIL dataset \cite{wang2019joint} comprises six distinct gestures (namely, hand up, hand down, hand left, hand right, hand circle, and hand cross), executed by a single user at 16 distinct locations within a confined space. Notably, the ARIL dataset, while featuring only one environmental variable (location), incorporates the use of universal software radio peripheral (USRP) devices for CSI data collection. We employ this dataset to assess the versatility of WiOpen under different openness $\mathcal{P}$ across different wireless devices. Otherwise, we also use ARIL to evaluate the cross-domain performance of WiOpen. The ARIL dataset encompasses 1392 samples totally. 

\subsection{Implementation Details}

In our implementation, the data acquisition and preprocessing were realized using Matlab, whereas the OSGR network was constructed using the PyTorch framework. Following extensive experimentation, we determined the optimal hyperparameters, which were set as follows: $\xi = 2$, $\lambda = 1$, and $\gamma = 0.05$. The number of neighbors, $K$, used for selecting test sample labels was set to $50$. During training, we utilized an initial learning rate of $0.001$, which was reduced by a factor of $10$ every $15$ epochs for a total of $50$ epochs. We employed the Adam optimizer for our model, and the feature extraction backbone network, $f_{\theta1}(x)$, is a ResNet18 network, while the $f_{\theta2}(x)$ is a 7- layer CNN. The code is available at \url{https://github.com/purpleleaves007/WiOpen}.

\begin{table}
  \centering
  \caption{AUROC comparison between different methods on Widar3.0. The best performance values are highlighted in bold.}
  \begin{tabular}{@{}cccccc@{}}
    \toprule
    Method $\backslash$ $O_P$ & 0.11 & 0.16 & 0.22 & 0.29 & 0.40 \\
    \midrule
    Softmax \cite{bendale2016towards} & 0.67 & 0.76 & 0.73 & 0.74 & 0.71\\
    WiGRUNT \cite{gu2022wigrunt} & 0.69 & 0.76 & 0.73 & 0.75 & 0.72\\
    Wione \cite{gu2021wione} & 0.68 & 0.75 & 0.75 & 0.76 & 0.69\\
    WiOpen & \textbf{0.74} & \textbf{0.80} & \textbf{0.81} & \textbf{0.79} & \textbf{0.76}\\
    \bottomrule
  \end{tabular}
  \label{tab:aurocw}
\end{table}

\subsection{Overall Performance}
\label{sec:ove}
In this study, we have chosen to assess OSGR performance using close-set accuracy and the area under the Receiver Operating Characteristic curve (AUROC). Close-set accuracy focuses on evaluating the classification performance with respect to known classes, while AUROC \cite{huang2022class} serves as a robust measure for assessing the ability to distinguish unknown classes. An AUROC value of "1" indicates complete separability between known and unknown classes. We also conducted a comparative analysis between WiOpen and three reference models: a baseline model based on Softmax, a SOTA Wi-Fi gesture recognition system WiGRUNT \cite{gu2022wigrunt} and a prototype based method Wione \cite{gu2021wione}. 


\begin{table}
  \centering
  \caption{Close-set accuracy comparison between different methods on Widar3.0. The best performance values are highlighted in bold.}
    \begin{tabular}{@{}cccccc@{}}
    \toprule
    Method $\backslash$ $O_P$ & 0.11 & 0.16 & 0.22 & 0.29 & 0.40 \\
    \midrule
    Softmax \cite{bendale2016towards} & 96.27\% & 95.11\% & 95.11\% & 95.56\% & 95.33\%\\
    WiGRUNT \cite{gu2022wigrunt} & \textbf{96.81\%} & 95.22\% & 96.11\% & 95.85\% & 96.89\%\\
    Wione \cite{gu2021wione} & 96.30\% & 95.29\% & 96.22\% & 96.15\% & 96.22\%\\
    WiOpen & 96.59\% & \textbf{95.82\%} & \textbf{96.56\%} & \textbf{96.59\%} & \textbf{97.11\%}\\
    \bottomrule
  \end{tabular}
  \label{tab:claccw}
\end{table}

\textbf{Results on Widar3.0.} To evaluate the performance of the WiOpen in open-set scenarios, we conducted experiments using 10,125 samples from the first environment and performed testing at different openness levels ($\mathcal{P} = 0.11-0.40$). When $\mathcal{P} = 0.29$, it indicates that our model was trained using 2,700 samples from three known classes. Within the test set, we allocated 675 samples from these three known classes as "known" samples, while the remaining 6,750 samples from six other classes were considered "unknown." This setup mirrors a real-world scenario where a substantial number of unknown samples are encountered. For AUROC calculations, we assigned label "0" to known classes and label "1" to unknown classes.

\begin{table}
  \centering
  \caption{AUROC comparison between different methods on ARIL. The best performance values are highlighted in bold.}
  \begin{tabular}{@{}cccccc@{}}
    \toprule
    Method $\backslash$ $O_P$ & 0.05 & 0.11 & 0.18 & 0.29 \\
    \midrule
    Softmax \cite{bendale2016towards} & 0.64 & 0.65 & 0.59 & 0.71\\
    WiGRUNT \cite{gu2022wigrunt} & 0.65 & 0.66 & 0.58 &0.72\\
    Wione \cite{gu2021wione} & 0.66 & 0.66 & 0.57 &0.70\\
    WiOpen & \textbf{0.73} & \textbf{0.67} & \textbf{0.72} &\textbf{0.76}\\
    \bottomrule
  \end{tabular}
  \label{tab:auroca}
\end{table}

\begin{figure}
    \centering
    \includegraphics[width=0.5\linewidth]{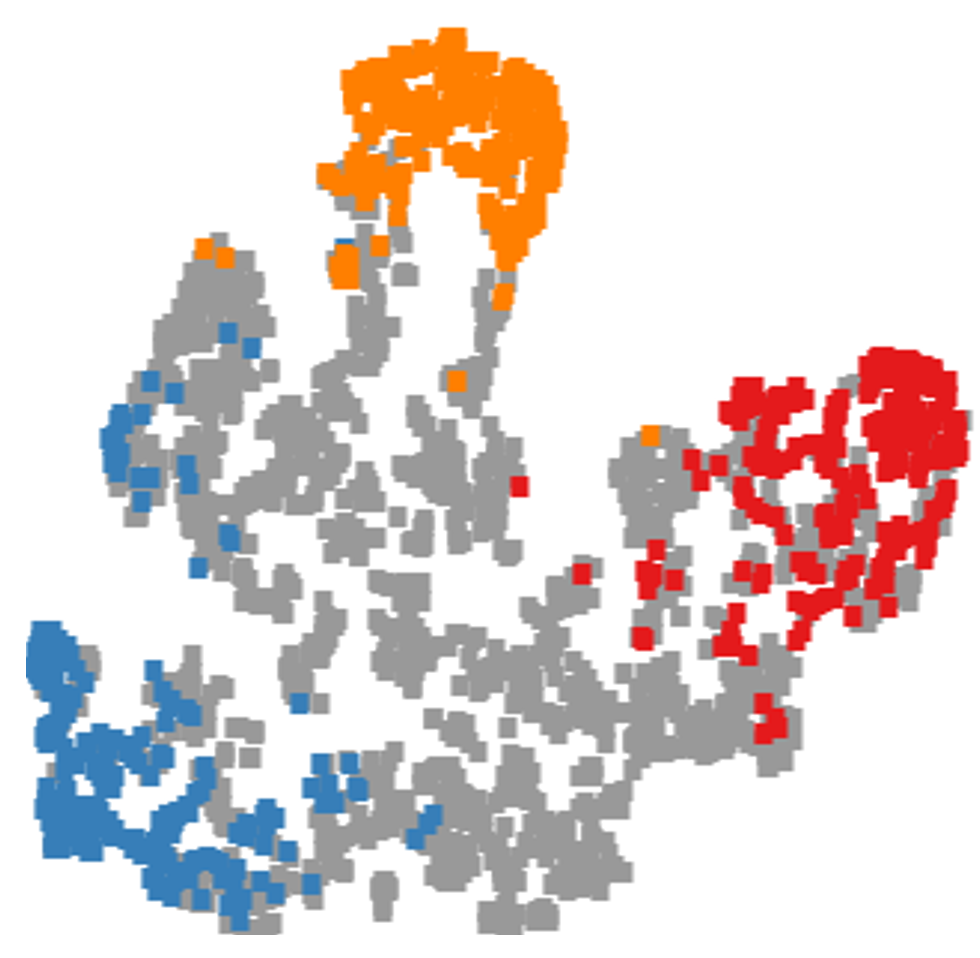}
    \caption{The distribution of samples in the feature space.}
    \label{fig:dunwop}
\end{figure}

The experimental results, as outlined in Tables \ref{tab:aurocw} and \ref{tab:claccw}, underscore the superior performance of WiOpen in open-set scenarios compared to traditional methods. WiOpen not only maintains competitive recognition accuracy but also excels in identifying unknown samples, demonstrating a consistently higher AUROC ranging from $0.03$ to $0.07$ compared to other methods. Notably, the transition from conventional gesture recognition systems, such as WiGRUNT, to OSGR systems often involves a complex search process to determine the optimal score threshold. In contrast, WiOpen dynamically adapts the threshold during the training process. WiOpen's training and unknown rejection strategies, grounded in neighborhood structures, offer an improvement over prototype-based methods like Wione. As evident in Tables \ref{tab:aurocw} and \ref{tab:claccw}, WiOpen achieves a balanced performance in recognizing known samples and rejecting unknown samples. It's crucial to note that when $\mathcal{P} = 0.11$, the AUROC for all approaches is notably low. This phenomenon is attributed to the high similarity between the sixth and seventh gestures (Zigzag and N), which can be considered essentially the same gesture rotated in space. This emphasizes the importance of addressing the distinction between intra-set and extra-set gestures, especially when they exhibit varying levels of similarity in open-set scenarios. This presents an intriguing avenue for future research. To visually represent the impact of the OSGR network, we plotted the distribution of samples in the feature space after processing, as depicted in Figure \ref{fig:dunwop}. The figure illustrates that the OSGR network effectively mitigates a portion of domain uncertainty compared to the state depicted in Figure \ref{dunm}.

\begin{table}
  \centering
  \caption{Close-set accuracy comparison between different methods on ARIL. The best performance values are highlighted in bold.}
    \begin{tabular}{@{}cccccc@{}}
    \toprule
    Method $\backslash$ $O_P$ & 0.05 & 0.11 & 0.18 & 0.29 \\
    \midrule
    Softmax \cite{bendale2016towards} & 86.25\% & 88.02\% & 93.06\% & 91.67\%\\
    WiGRUNT \cite{gu2022wigrunt} & 87.50\% & 88.54\% & 93.75\% & 92.71\%\\
    Wione \cite{gu2021wione} & 87.95\% & 88.79\% & 92.96\% & 90.52\%\\
    WiOpen & \textbf{89.58\%} & \textbf{93.23\%} & \textbf{93.75\%} &\textbf{94.79\%}\\
    \bottomrule
  \end{tabular}
  \label{tab:clacca}
\end{table}

\begin{table}
  \centering
  \caption{Close-set accuracy comparison between different methods on ARIL. The best performance values are highlighted in bold.}
    \begin{tabular}{@{}cccccc@{}}
    \toprule
    {\multirow{2}{*}{Method}} & \multicolumn{2}{c}{Cross-Ori} & \multicolumn{2}{c}{Cross-Loc}\\
    \multicolumn{2}{c}{} AUROC & Acc & AUROC & Acc &\\
    \midrule
    Softmax \cite{bendale2016towards} & 0.70 & 90.80\% & 0.76 & 97.19\%\\
    WiGRUNT \cite{gu2022wigrunt} & 0.72 & 91.70\% & 0.76 & \textbf{97.33\%}\\
    Wione \cite{gu2021wione} & 0.69 & 90.80\% & 0.74 & 96.87\%\\
    WiOpen & \textbf{0.78} & \textbf{92.59\%} & \textbf{0.81} &97.19\%\\
    \bottomrule
  \end{tabular}
  \label{tab:cdops}
\end{table}

\textbf{Results on ARIL.} For ARIL, we use all 1,392 samples to evaluate WiOpen. Testing was conducted across a range of openness levels ($\mathcal{P} = 0.05-0.29$). The parameters and experimental procedures remained consistent with those employed for the WiDar3.0 dataset. It's worth noting that when evaluating the softmax and WiGRUNT systems, a new search for the optimal threshold was required, introducing notable inconvenience and limitations.

The experimental results, presented in Tables 4 and 5, firmly establish WiOpen's superiority in both recognizing known samples and effectively rejecting unknown samples. These results underscore WiOpen's robust performance.

\textbf{Results on Open-Set + Cross Domain.} To ascertain WiOpen's performance in more demanding environments, we combined open-set scenarios with cross-domain scenarios. Specifically, we conducted experiments utilizing data from the 1st environment of the WiDar3.0 dataset, applying open-set setting across different locations and orientations, all under an openness level of $0.29$.

\begin{table*}[htp]
  \centering
  \caption{Cross domain gesture recognition results compared with state-of-the-art solutions.(I-D, C-L, C-O, C-E and C-U means In-Domain, Cross Location, Orientation, Environment and User, respectively. 6D and 1D means use 6 and 1 pairs of transmitter-receivers, respectively).}
    \begin{tabular}{@{}cccccccccccccc@{}}
    \toprule
    {\multirow{2}{*}{Method}} & \multirow{2}{*}{Pub} & \multirow{2}{*}{Year} & \multicolumn{5}{c}{WiDar3.0-6D} & \multicolumn{4}{c}{WiDar3.0-1D}& \multicolumn{1}{c}{ARIL}\\
    \multicolumn{2}{c}{} & & I-D & C-L & C-O & C-E & C-U & C-E & C-L & C-O & C-U & C-L & \\
    \midrule
    WiDar3.0 \cite{zhang2021widar3} & IEEE TPAMI & 2021 & 92.7\% & 89.7\% & 82.6\% & 92.4\% & - & - & - & - & - & -\\
    WiHF \cite{li2020wihf} & IEEE TMC & 2020 & 97.65\% & 92.07\% & 82.38\% & 89.67\% & - & - & - & - & - & -\\
    WiGRUNT \cite{gu2022wigrunt} & IEEE THMS & 2022 & \underline{99.71\%} & 96.62\% & 93.85\% & 93.73\% & - & - & - & - & - & -\\
    PAC-CSI \cite{su2023real} & IEEE JSAC & 2023 & 99.46\% & \underline{98.77\%} & \textbf{98.90\%} & \underline{96.47\%} & \underline{97.54\%} & - & - & - & - & -\\
    SelfReg \cite{kim2021selfreg} & IEEE ICCV & 2021 & - & - & - & - & - & 39.11\% & 76.71\% & 86.67\% & 53.10\% & 44.45\% \\
    WiSGP \cite{liu2023generalizing} & IEEE TMC & 2023 & - & - & - & - & - & \underline{43.17\%} & \underline{78.49\%} & \underline{88.46\%} & \underline{56.77\%} & \underline{48.74\%}\\
    WiSR \cite{liu2023wisr} & IEEE TMC & 2023 & - & - & - & - & - & 42.52\% & 77.51\% & \textbf{88.80\%} & 55.18\% & 48.64\%\\
    WiOpen & - & 2023 & \textbf{99.78\%} & \textbf{98.81\%} & \underline{98.05\%} & \textbf{97.99\%} & \textbf{98.47\%} & \textbf{84.44\%} & \textbf{86.40\%} & 77.67\% & \textbf{82.71\%} & \textbf{73.61\%}\\
    \bottomrule
  \end{tabular}
  \label{tab:crope}
\end{table*}

The experimental results, as detailed in Table \ref{tab:cdops}, affirm that WiOpen maintains its proficiency in recognizing known samples and effectively rejecting unknown samples, even in cross-domain open-set scenarios. Notably, the introduction of domain discrepancies poses additional challenges. The successful performance of WiOpen in cross-domain open-set recognition is especially promising, as it closely aligns with real-world applications and merits further exploration.


\subsection{Cross Domain Performance}
\label{sec:cro}
To facilitate a more comprehensive comparison of WiOpen with other SOTA Wi-Fi based gesture recognition systems and to underscore the efficacy of our proposed uncertainty reduction method, we extended our experiments into cross-domain tasks. To ensure compatibility with established solutions, we conducted performance evaluations using the WiDar3.0 dataset in two distinct scenarios: one involving all six pairs of transmitter-receivers and the other employing only a single pair of transmitter-receiver. In the case of utilizing all six pairs of devices, our dataset settings closely align with the SOTA PAC-CSI \cite{su2023real}. It's noteworthy that they opted to select 80\% of the testing samples for the test set, a distinction from our approach in which we did not perform any sample selection for the test set. In the scenario involving just one pair of transmitter-receiver, our dataset settings mirror those of WiSR \cite{liu2023wisr}. Furthermore, we also conducted experiments on the ARIL dataset to evaluate the performance of WiOpen in a broader context.


\begin{figure}
    \centering
    \includegraphics[width=0.9\linewidth]{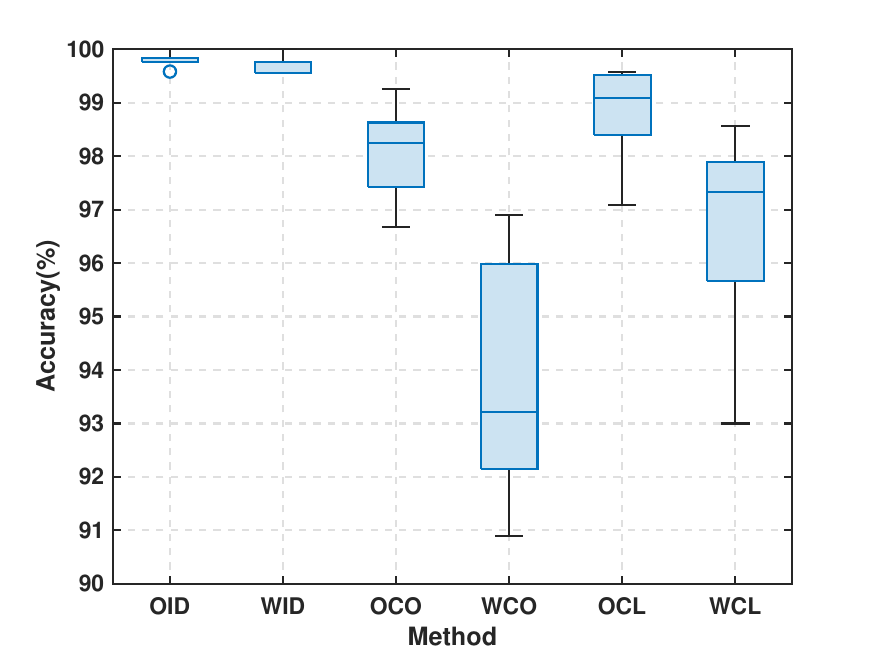}
    \caption{Box-plot of WiOpen and WiGRUNT, ID, CL and CO indicates in-domain, cross-location and cross-orientation, respectively, and O and W represents WiOpen and WiGRUNT, OID means in-domain results with WiOpen method.}
    \label{fig:boxfigure}
    \vspace{-0.2in}
\end{figure}

The experimental results are summarized in Table \ref{tab:crope}. In the WiDar3.0-6D setting, WiOpen exhibits slightly lower performance than the state-of-the-art PCA-CSI \cite{su2023real} in the cross-orientation task. However, it outperforms both WiGRUNT and PCA-CSI in other settings, showcasing superior overall performance. In the WiDar3.0-1D and ARIL scenarios, WiOpen lags behind WiSGP and WiSR in the cross-orientation task. Nevertheless, it significantly outperforms WiSGP \cite{liu2023generalizing} and WiSR \cite{liu2023wisr} in other tasks. Particularly noteworthy is WiOpen's substantial lead of over 41\% in cross-environment scenarios. To further assess the system's robustness, we conducted a comparative analysis between WiGRUNT \cite{gu2022wigrunt} and WiOpen under the WiDar3.0-6D setting. The results are depicted in Figure \ref{fig:boxfigure}. Clearly, WiOpen not only achieves higher accuracy but also demonstrates less performance variance across different domains compared to WiGRUNT. These findings underscore the superior robustness of WiOpen.


The key to WiOpen's success in cross-domain tasks lies in its innovative approach to uncertainty elimination. While initially designed to address open-set challenges in Wi-Fi gesture recognition, the domain uncertainty reduction facilitated by the OSGR network proves to be advantageous for cross-domain applications. The elimination of domain uncertainty induces the convergence of feature spaces for individual classes. The features acquired through OSGR learning remove domain-specific irregularities, fundamentally alleviating domain disparities and augmenting domain generalization capabilities. This further underscores the effectiveness of WiOpen, highlighting the central role played by its foundational principle of uncertainty reduction.

\subsection{Sensitivity Analysis}

In this section, we conduct sensitivity analyses on WiOpen to assess the impact of various factors on its performance.

\textbf{Impact of Threshold.}

\begin{figure}
    \centering
    \includegraphics[width=0.9\linewidth]{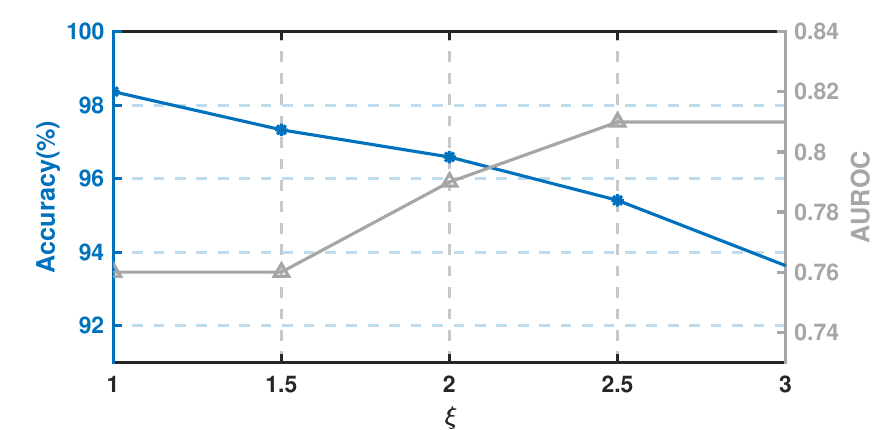}
    \caption{Impact of Threshold $\xi$.}
    \label{fig:thr}
\end{figure}

To evaluate the sensitivity of WiOpen to the threshold $\xi$, we conducted experiments with $\xi$ values ranging from 1 to 3 while keeping $\mathcal{P}=0.29$. The results, displayed in Figure \ref{fig:thr}, demonstrate a clear relationship between the threshold and system performance. As the threshold increases, the close-set accuracy decreases, but the system's effectiveness in rejecting unknown samples improves. To strike a balance between these performance aspects, we selected $\xi=2$.

\textbf{Impact of Construction Loss.}

\begin{figure}[t]
    \centering
    \includegraphics[width=0.9\linewidth]{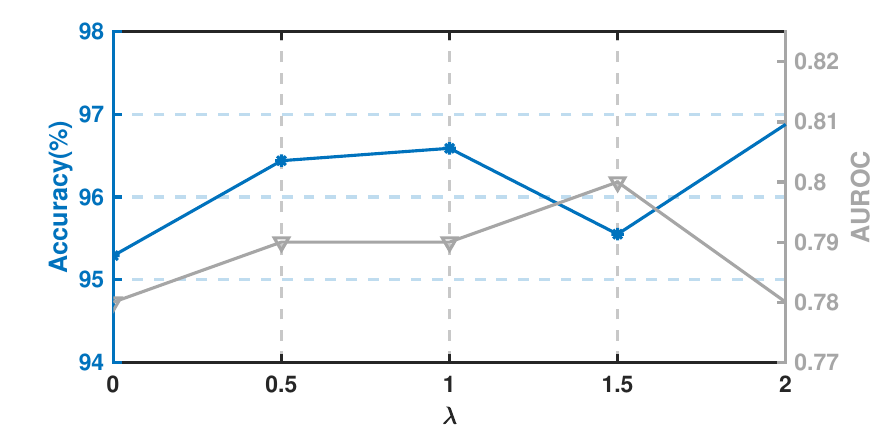}
    \caption{Impact of Construction Loss.}
    \label{fig:conloss}
\end{figure}

To analyze the influence of the construction loss, we performed experiments with $\lambda$ values ranging from 0 to 2 while maintaining $\mathcal{P}=0.29$. The results, as illustrated in Figure \ref{fig:conloss}, show that the construction loss significantly contributes to WiOpen's ability to acquire valuable knowledge. In our experiments, we opted for $\lambda=1$.

\textbf{Impact of Different Amounts of Training Data.}

\begin{figure}
    \centering
    \includegraphics[width=0.9\linewidth]{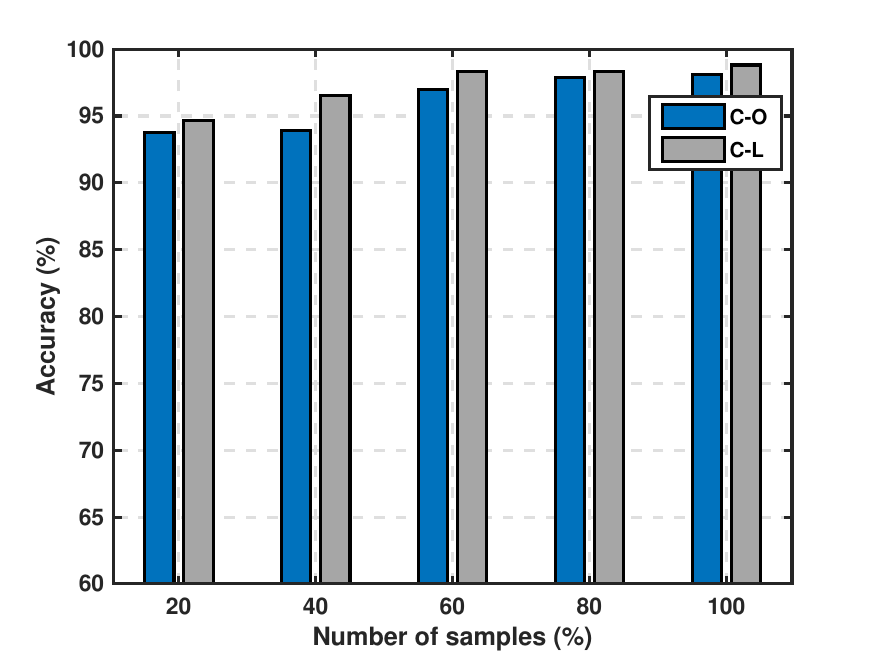}
    \caption{Impact of Different Amounts of Training Data.}
    \label{fig:numsam}
\end{figure}

To evaluate WiOpen's capability to learn classification knowledge with limited training samples, we conducted experiments by reducing the size of the training set to 20\% - 80\% of the original training set. The results, depicted in Figure \ref{fig:numsam}, reveal that even with a reduced sample size, WiOpen continues to perform well in cross-location and cross-direction domain recognition tasks. This underscores WiOpen's aptitude for acquiring classification knowledge.

\section{Conclusions}

In this paper, we have introduced WiOpen, a pioneering Wi-Fi-based OSGR system. We commenced our study with a comprehensive analysis of open-set challenges within the realm of Wi-Fi sensing, shedding light on the intrinsic correlation between Wi-Fi-based OSGR and uncertainty. Building upon this correlation analysis, we have put forth solutions designed to eradicate uncertainty in Wi-Fi sensing and establish decision boundaries. Our experimental results not only underscore the remarkable effectiveness of WiOpen in open-set gesture recognition, but also validate the profound advantages of WiOpen in cross-domain and small-sample recognition tasks. This substantiates WiOpen's unique ability to glean meaningful knowledge, making it a robust system for real-world applications. For future works, we will continue to address the intricate challenges posed by the high similarity between unknown and known gestures in open-set scenarios. Furthermore, we aim to tackle the complexities introduced when the target domain contains unknown samples during domain adaptation, further enhancing their suitability for real-world applications.

\bibliographystyle{IEEEtran}
\bibliography{wiopen}

\begin{thebibliography}{10}
\providecommand{\url}[1]{#1}
\csname url@samestyle\endcsname
\providecommand{\newblock}{\relax}
\providecommand{\bibinfo}[2]{#2}
\providecommand{\BIBentrySTDinterwordspacing}{\spaceskip=0pt\relax}
\providecommand{\BIBentryALTinterwordstretchfactor}{4}
\providecommand{\BIBentryALTinterwordspacing}{\spaceskip=\fontdimen2\font plus
\BIBentryALTinterwordstretchfactor\fontdimen3\font minus
  \fontdimen4\font\relax}
\providecommand{\BIBforeignlanguage}[2]{{%
\expandafter\ifx\csname l@#1\endcsname\relax
\typeout{** WARNING: IEEEtran.bst: No hyphenation pattern has been}%
\typeout{** loaded for the language `#1'. Using the pattern for}%
\typeout{** the default language instead.}%
\else
\language=\csname l@#1\endcsname
\fi
#2}}
\providecommand{\BIBdecl}{\relax}
\BIBdecl

\bibitem{gu2022wigrunt}
Y.~Gu, X.~Zhang, Y.~Wang, M.~Wang, H.~Yan, Y.~Ji, Z.~Liu, J.~Li, and M.~Dong,
  ``Wigrunt: Wifi-enabled gesture recognition using dual-attention network,''
  \emph{IEEE Transactions on Human-Machine Systems}, vol.~52, no.~4, pp.
  736--746, 2022.

\bibitem{zhang2021widar3}
Y.~Zhang, Y.~Zheng, K.~Qian, G.~Zhang, Y.~Liu, C.~Wu, and Z.~Yang, ``Widar3. 0:
  Zero-effort cross-domain gesture recognition with wi-fi,'' \emph{IEEE
  Transactions on Pattern Analysis and Machine Intelligence}, vol.~44, no.~11,
  pp. 8671--8688, 2021.

\bibitem{li2020wihf}
C.~Li, M.~Liu, and Z.~Cao, ``Wihf: Gesture and user recognition with wifi,''
  \emph{IEEE Transactions on Mobile Computing}, vol.~21, no.~2, pp. 757--768,
  2020.

\bibitem{liu2023wisr}
S.~Liu, Z.~Chen, M.~Wu, C.~Liu, and L.~Chen, ``Wisr: Wireless domain
  generalization based on style randomization,'' \emph{IEEE Transactions on
  Mobile Computing}, 2023.

\bibitem{gu2021wione}
Y.~Gu, H.~Yan, M.~Dong, M.~Wang, X.~Zhang, Z.~Liu, and F.~Ren, ``Wione:
  One-shot learning for environment-robust device-free user authentication via
  commodity wi-fi in man--machine system,'' \emph{IEEE Transactions on
  Computational Social Systems}, vol.~8, no.~3, pp. 630--642, 2021.

\bibitem{wang2022caution}
D.~Wang, J.~Yang, W.~Cui, L.~Xie, and S.~Sun, ``Caution: A robust wifi-based
  human authentication system via few-shot open-set recognition,'' \emph{IEEE
  Internet of Things Journal}, vol.~9, no.~18, pp. 17\,323--17\,333, 2022.

\bibitem{tan2020enabling}
S.~Tan, J.~Yang, and Y.~Chen, ``Enabling fine-grained finger gesture
  recognition on commodity wifi devices,'' \emph{IEEE Transactions on Mobile
  Computing}, vol.~21, no.~8, pp. 2789--2802, 2020.

\bibitem{tan2019multitrack}
S.~Tan, L.~Zhang, Z.~Wang, and J.~Yang, ``Multitrack: Multi-user tracking and
  activity recognition using commodity wifi,'' in \emph{Proceedings of the 2019
  CHI Conference on Human Factors in Computing Systems}, 2019, pp. 1--12.

\bibitem{wang2016human}
H.~Wang, D.~Zhang, J.~Ma, Y.~Wang, Y.~Wang, D.~Wu, T.~Gu, and B.~Xie, ``Human
  respiration detection with commodity wifi devices: Do user location and body
  orientation matter?'' in \emph{Proceedings of the 2016 ACM international
  joint conference on pervasive and ubiquitous computing}, 2016, pp. 25--36.

\bibitem{niu2018fresnel}
K.~Niu, F.~Zhang, Z.~Chang, and D.~Zhang, ``A fresnel diffraction model based
  human respiration detection system using cots wi-fi devices,'' in
  \emph{Proceedings of the 2018 ACM international joint conference and 2018
  international symposium on pervasive and ubiquitous computing and wearable
  computers}, 2018, pp. 416--419.

\bibitem{pu2013whole}
Q.~Pu, S.~Gupta, S.~Gollakota, and S.~Patel, ``Whole-home gesture recognition
  using wireless signals,'' in \emph{Proceedings of the 19th annual
  international conference on Mobile computing \& networking}, 2013, pp.
  27--38.

\bibitem{wang2014eyes}
Y.~Wang, J.~Liu, Y.~Chen, M.~Gruteser, J.~Yang, and H.~Liu, ``E-eyes:
  device-free location-oriented activity identification using fine-grained wifi
  signatures,'' in \emph{Proceedings of the 20th annual international
  conference on Mobile computing and networking}, 2014, pp. 617--628.

\bibitem{abdelnasser2015wigest}
H.~Abdelnasser, M.~Youssef, and K.~A. Harras, ``Wigest: A ubiquitous wifi-based
  gesture recognition system,'' in \emph{2015 IEEE conference on computer
  communications (INFOCOM)}.\hskip 1em plus 0.5em minus 0.4em\relax IEEE, 2015,
  pp. 1472--1480.

\bibitem{venkatnarayan2018multi}
R.~H. Venkatnarayan, G.~Page, and M.~Shahzad, ``Multi-user gesture recognition
  using wifi,'' in \emph{Proceedings of the 16th Annual International
  Conference on Mobile Systems, Applications, and Services}, 2018, pp.
  401--413.

\bibitem{sun2015widraw}
L.~Sun, S.~Sen, D.~Koutsonikolas, and K.-H. Kim, ``Widraw: Enabling hands-free
  drawing in the air on commodity wifi devices,'' in \emph{Proceedings of the
  21st Annual International Conference on Mobile Computing and Networking},
  2015, pp. 77--89.

\bibitem{yu2018qgesture}
N.~Yu, W.~Wang, A.~X. Liu, and L.~Kong, ``Qgesture: Quantifying gesture
  distance and direction with wifi signals,'' \emph{Proceedings of the ACM on
  Interactive, Mobile, Wearable and Ubiquitous Technologies}, vol.~2, no.~1,
  pp. 1--23, 2018.

\bibitem{li2016wifinger}
H.~Li, W.~Yang, J.~Wang, Y.~Xu, and L.~Huang, ``Wifinger: talk to your smart
  devices with finger-grained gesture,'' in \emph{Proceedings of the 2016 ACM
  International Joint Conference on Pervasive and Ubiquitous Computing}, 2016,
  pp. 250--261.

\bibitem{ali2015keystroke}
K.~Ali, A.~X. Liu, W.~Wang, and M.~Shahzad, ``Keystroke recognition using wifi
  signals,'' in \emph{Proceedings of the 21st annual international conference
  on mobile computing and networking}, 2015, pp. 90--102.

\bibitem{zhang2020wisign}
L.~Zhang, Y.~Zhang, and X.~Zheng, ``Wisign: Ubiquitous american sign language
  recognition using commercial wi-fi devices,'' \emph{ACM Transactions on
  Intelligent Systems and Technology (TIST)}, vol.~11, no.~3, pp. 1--24, 2020.

\bibitem{zheng2019zero}
Y.~Zheng, Y.~Zhang, K.~Qian, G.~Zhang, Y.~Liu, C.~Wu, and Z.~Yang,
  ``Zero-effort cross-domain gesture recognition with wi-fi,'' in
  \emph{Proceedings of the 17th Annual International Conference on Mobile
  Systems, Applications, and Services}, 2019, pp. 313--325.

\bibitem{liu2023generalizing}
S.~Liu, Z.~Chen, M.~Wu, H.~Wang, B.~Xing, and L.~Chen, ``Generalizing wireless
  cross-multiple-factor gesture recognition to unseen domains,'' \emph{IEEE
  Transactions on Mobile Computing}, 2023.

\bibitem{su2023real}
J.~Su, Q.~Mao, Z.~Liao, Z.~Sheng, C.~Huang, and X.~Zhang, ``A real-time
  cross-domain wi-fi-based gesture recognition system for digital twins,''
  \emph{IEEE Journal on Selected Areas in Communications}, 2023.

\bibitem{gu2023attention}
Y.~Gu, H.~Yan, X.~Zhang, Y.~Wang, J.~Huang, Y.~Ji, and F.~Ren,
  ``Attention-based gesture recognition using commodity wifi devices,''
  \emph{IEEE Sensors Journal}, vol.~23, no.~9, pp. 9685--9696, 2023.

\bibitem{feng2022wi}
C.~Feng, N.~Wang, Y.~Jiang, X.~Zheng, K.~Li, Z.~Wang, and X.~Chen,
  ``Wi-learner: Towards one-shot learning for cross-domain wi-fi based gesture
  recognition,'' \emph{Proceedings of the ACM on Interactive, Mobile, Wearable
  and Ubiquitous Technologies}, vol.~6, no.~3, pp. 1--27, 2022.

\bibitem{bendale2015towards}
A.~Bendale and T.~Boult, ``Towards open world recognition,'' in
  \emph{Proceedings of the IEEE conference on computer vision and pattern
  recognition}, 2015, pp. 1893--1902.

\bibitem{mendes2017nearest}
P.~R. Mendes~J{\'u}nior, R.~M. De~Souza, R.~d.~O. Werneck, B.~V. Stein, D.~V.
  Pazinato, W.~R. de~Almeida, O.~A. Penatti, R.~d.~S. Torres, and A.~Rocha,
  ``Nearest neighbors distance ratio open-set classifier,'' \emph{Machine
  Learning}, vol. 106, no.~3, pp. 359--386, 2017.

\bibitem{scheirer2012toward}
W.~J. Scheirer, A.~de~Rezende~Rocha, A.~Sapkota, and T.~E. Boult, ``Toward open
  set recognition,'' \emph{IEEE transactions on pattern analysis and machine
  intelligence}, vol.~35, no.~7, pp. 1757--1772, 2012.

\bibitem{zhou2021learning}
D.-W. Zhou, H.-J. Ye, and D.-C. Zhan, ``Learning placeholders for open-set
  recognition,'' in \emph{Proceedings of the IEEE/CVF conference on computer
  vision and pattern recognition}, 2021, pp. 4401--4410.

\bibitem{bendale2016towards}
A.~Bendale and T.~E. Boult, ``Towards open set deep networks,'' in
  \emph{Proceedings of the IEEE conference on computer vision and pattern
  recognition}, 2016, pp. 1563--1572.

\bibitem{hendrycks2016baseline}
D.~Hendrycks and K.~Gimpel, ``A baseline for detecting misclassified and
  out-of-distribution examples in neural networks,'' in \emph{International
  Conference on Learning Representations}, 2016.

\bibitem{sun2020conditional}
X.~Sun, Z.~Yang, C.~Zhang, K.-V. Ling, and G.~Peng, ``Conditional gaussian
  distribution learning for open set recognition,'' in \emph{Proceedings of the
  IEEE/CVF Conference on Computer Vision and Pattern Recognition}, 2020, pp.
  13\,480--13\,489.

\bibitem{yang2020convolutional}
H.-M. Yang, X.-Y. Zhang, F.~Yin, Q.~Yang, and C.-L. Liu, ``Convolutional
  prototype network for open set recognition,'' \emph{IEEE Transactions on
  Pattern Analysis and Machine Intelligence}, vol.~44, no.~5, pp. 2358--2370,
  2020.

\bibitem{chen2020learning}
G.~Chen, L.~Qiao, Y.~Shi, P.~Peng, J.~Li, T.~Huang, S.~Pu, and Y.~Tian,
  ``Learning open set network with discriminative reciprocal points,'' in
  \emph{Computer Vision--ECCV 2020: 16th European Conference, Glasgow, UK,
  August 23--28, 2020, Proceedings, Part III 16}.\hskip 1em plus 0.5em minus
  0.4em\relax Springer, 2020, pp. 507--522.

\bibitem{chen2021adversarial}
G.~Chen, P.~Peng, X.~Wang, and Y.~Tian, ``Adversarial reciprocal points
  learning for open set recognition,'' \emph{IEEE Transactions on Pattern
  Analysis and Machine Intelligence}, vol.~44, no.~11, pp. 8065--8081, 2021.

\bibitem{huang2022class}
H.~Huang, Y.~Wang, Q.~Hu, and M.-M. Cheng, ``Class-specific semantic
  reconstruction for open set recognition,'' \emph{IEEE transactions on pattern
  analysis and machine intelligence}, vol.~45, no.~4, pp. 4214--4228, 2022.

\bibitem{geng2020recent}
C.~Geng, S.-j. Huang, and S.~Chen, ``Recent advances in open set recognition: A
  survey,'' \emph{IEEE transactions on pattern analysis and machine
  intelligence}, vol.~43, no.~10, pp. 3614--3631, 2020.

\bibitem{huang2021phaseanti}
J.~Huang, B.~Liu, C.~Miao, Y.~Lu, Q.~Zheng, Y.~Wu, J.~Liu, L.~Su, and C.~W.
  Chen, ``Phaseanti: An anti-interference wifi-based activity recognition
  system using interference-independent phase component,'' \emph{IEEE
  Transactions on Mobile Computing}, vol.~22, no.~5, pp. 2938--2954, 2023.

\bibitem{liu2021adaptive}
J.~Liu, H.~Xu, L.~Wang, Y.~Xu, C.~Qian, J.~Huang, and H.~Huang, ``Adaptive
  asynchronous federated learning in resource-constrained edge computing,''
  \emph{IEEE Transactions on Mobile Computing}, vol.~22, no.~2, pp. 674--690,
  2023.

\bibitem{yan2019wiact}
H.~Yan, Y.~Zhang, Y.~Wang, and K.~Xu, ``Wiact: A passive wifi-based human
  activity recognition system,'' \emph{IEEE Sensors Journal}, vol.~20, no.~1,
  pp. 296--305, 2019.

\bibitem{zhang2023wital}
X.~Zhang, Y.~Gu, H.~Yan, Y.~Wang, M.~Dong, K.~Ota, F.~Ren, and Y.~Ji, ``Wital:
  A cots wifi devices based vital signs monitoring system using nlos sensing
  model,'' \emph{IEEE Transactions on Human-Machine Systems}, vol.~53, no.~3,
  pp. 629--641, 2023.

\bibitem{zhang2019towards}
F.~Zhang, K.~Niu, J.~Xiong, B.~Jin, T.~Gu, Y.~Jiang, and D.~Zhang, ``Towards a
  diffraction-based sensing approach on human activity recognition,''
  \emph{Proceedings of the ACM on Interactive, Mobile, Wearable and Ubiquitous
  Technologies}, vol.~3, no.~1, pp. 1--25, 2019.

\bibitem{scheirer2014probability}
W.~J. Scheirer, L.~P. Jain, and T.~E. Boult, ``Probability models for open set
  recognition,'' \emph{IEEE transactions on pattern analysis and machine
  intelligence}, vol.~36, no.~11, pp. 2317--2324, 2014.

\bibitem{wu2020fingerdraw}
D.~Wu, R.~Gao, Y.~Zeng, J.~Liu, L.~Wang, T.~Gu, and D.~Zhang, ``Fingerdraw:
  Sub-wavelength level finger motion tracking with wifi signals,''
  \emph{Proceedings of the ACM on Interactive, Mobile, Wearable and Ubiquitous
  Technologies}, vol.~4, no.~1, pp. 1--27, 2020.

\bibitem{zeng2019farsense}
Y.~Zeng, D.~Wu, J.~Xiong, E.~Yi, R.~Gao, and D.~Zhang, ``Farsense: Pushing the
  range limit of wifi-based respiration sensing with csi ratio of two
  antennas,'' \emph{Proceedings of the ACM on Interactive, Mobile, Wearable and
  Ubiquitous Technologies}, vol.~3, no.~3, pp. 1--26, 2019.

\bibitem{wu2018improving}
Z.~Wu, A.~A. Efros, and S.~X. Yu, ``Improving generalization via scalable
  neighborhood component analysis,'' in \emph{Proceedings of the european
  conference on computer vision (ECCV)}, 2018, pp. 685--701.

\bibitem{wang2019joint}
F.~Wang, J.~Feng, Y.~Zhao, X.~Zhang, S.~Zhang, and J.~Han, ``Joint activity
  recognition and indoor localization with wifi fingerprints,'' \emph{IEEE
  Access}, vol.~7, pp. 80\,058--80\,068, 2019.

\bibitem{kim2021selfreg}
D.~Kim, Y.~Yoo, S.~Park, J.~Kim, and J.~Lee, ``Selfreg: Self-supervised
  contrastive regularization for domain generalization,'' in \emph{Proceedings
  of the IEEE/CVF International Conference on Computer Vision}, 2021, pp.
  9619--9628.

\end{thebibliography}

\end{document}